\documentclass{aa}

\usepackage{graphicx}
\usepackage{float}
\usepackage{txfonts}
\usepackage{multirow}
\usepackage{amsmath}	
\usepackage{xcolor}
\usepackage{url}
\usepackage{booktabs}

\usepackage{pifont}
\newcommand{\cmark}{\ding{51}}%
\newcommand{\xmark}{\ding{55}}%

\usepackage[draft=false]{hyperref}

\defcitealias{Spetsieri_prep}{Spetsieri~et~al.~in prep.}

\begin{document}


\title{Reeling in the Whirlpool: the distance to M~51 clarified by Cepheids and the Type IIP SN~2005cs}

\titlerunning{Independent and precise distances to M51}

\author{ G. Cs\"ornyei
		\inst{1,2}
		\and
		R.~I. Anderson
		\inst{3}
		\and
 		C. Vogl
 		\inst{1,2,4}
 		\and
 		S. Taubenberger
 		\inst{1,2}
 		\and
        S. Blondin
        \inst{5}
        \and
 		B. Leibundgut
  		\inst{6}
 		\and
 		W. Hillebrandt
 		\inst{1,4}}

\institute{Max-Planck-Institute for Astrophysics, Karl-Schwarzschild-Str. 1, 85741 Garching, Germany\\
			 \email{csogeza@mpa-garching.mpg.de}
		\and
			Technical University Munich, TUM Department of Phyiscs, James-Franck-Str. 1., 85741 Garching, Germany 	
		\and
			École Polytechnique Fédérale de Lausanne (EPFL), Observatoire de Sauverny, Institute of Physics, Laboratory of Astrophysics, 1290 Versoix, Switzerland
		\and
			Exzellenzcluster ORIGINS, Boltzmannstr. 2, 85748 Garching, Germany
        \and
            Aix-Marseille Univ, CNRS, CNES, LAM, 13388 Marseille, France
		\and
			European Southern Observatory, Karl-Schwarzschild-Str. 2, 85741 Garching, Germany	
}

\date{Received XXX}

\abstract
{Despite being one of the best-known galaxies, the distance to the Whirlpool Galaxy, M~51, is still debated. Current estimates range from 6.02 to 9.09 Mpc, and different methods yield discrepant results. No Cepheid distance has been published for M~51 to date.}
{We aim to estimate a more reliable distance to M~51 through two independent methods: Cepheid variables and their period-luminosity relation, and an augmented version of the expanding photosphere method (EPM) on the Type IIP supernova SN~2005cs which exploded in this galaxy.}
{For the Cepheid variables, we analyse a recently published \emph{Hubble Space Telescope} catalogue of stars in M~51. By applying light curve and colour-magnitude diagram-based filtering, we select a high-quality sample of M~51 Cepheids to estimate the distance through the period-luminosity relation. For SN~2005cs, an emulator-based spectral fitting technique is applied, which allows for the fast and reliable estimation of physical parameters of the supernova atmosphere. We augment the established framework of EPM with these spectral models to obtain a precise distance to M~51.}
{The two resulting distance estimates are $D_{\textrm{Cep}} =  7.59 \pm 0.30$ Mpc and $D_{\textrm{2005cs}} = 7.34 \pm 0.39 $ Mpc using the Cepheid period-luminosity relation and the spectral modelling of SN~2005cs respectively. This is the first published Cepheid distance for this galaxy. The obtained values are precise to 4-5\% and fully consistent within $1\sigma$ uncertainties. Given that these two estimates are completely independent, one may combine them for an even more precise estimate, which yields $D_{\textrm{M~51}} = 7.50 \pm 0.24$ Mpc (3.2\% uncertainty).}
{Our distance estimates are in agreement with most of the results obtained previously for M~51, while being more precise than the earlier counterparts. They are however significantly lower than the TRGB estimates, which are often adopted for the distance to this galaxy. The results highlight the importance of direct cross-checks between independent distance estimates for quantifying systematic uncertainties. Given the large discrepancy, this finding can also influence distance-sensitive studies and their discussion for objects within M~51, as well as the estimation of the Hubble constant through the Type IIP standardizable candle method, for which SN~2005cs is a calibrator object.}



\maketitle

\section{Introduction}
The Whirlpool Galaxy (or Messier 51, M~51) is one of the best-known extragalactic objects in the sky for professional and amateur astronomers. Despite its proximity and many observations conducted by generations of astronomers, the distance of this galaxy remains uncertain compared to the state-of-the-art and other well-known galaxies, with little agreement between the different methods (see, e.g., \citealt{McQuinn2016}). This uncertainty is a limiting factor for studies that use the distance as an input, such as spatially resolved or luminosity-dependent analyses, e.g. in the context of star formation \citep{Heyer2022}, X-ray pulsar brightness \citep{Rodriguez2020} or interstellar medium kinematics \citep{Jorge2020}. Moreover, a distance uncertainty for M~51 also affects the extragalactic distance scale, as M~51 is one of the few calibrator hosts for the Type II supernova standardizable candle method (SCM) through the underluminous Type IIP supernova SN~2005cs \citep{deJaeger2022}.

In recent years a multitude of methods has been used to constrain the distance of M~51, which yielded results in a large range of values: the Tully-Fischer method (\citealt{Tully1977}, resulting in distances in the range of 4.9--12.2 Mpc, e.g. \citealt{Tutui1997}), the expanding photosphere method (EPM, \citealt{Kirshner1974}) applied to Type II supernovae (6.02--8.40 Mpc, e.g. \citealt{Takats2006, Vinko2012}), the planetary nebula luminosity function (PNLF, \citealt{Feldmeier1997}, 7.62--8.4 Mpc, e.g. \cite{Ciardullo2002}), the surface brightness fluctuation (SBF, \cite{Tonry1988}, 7.31--7.83 Mpc, e.g. \citealt{Tonry2001, Ciardullo2002}) and the tip of the red giant branch (TRGB, \citealt{Anand2022}. 8.58--9.09 Mpc, \citealt{Tikhonov2015, McQuinn2016}), which is most frequently quoted as the distance to M51 (as discussed in Sect.~\ref{sec:discussion}). For a more complete review of these distances and their determination, see \cite{McQuinn2016}. Interestingly, no Cepheid distance has been determined for this galaxy to date, despite the large number of observations obtained by the \emph{Hubble Space Telescope} ($HST$) in the past decades owing to, for example, several supernovae (see Table~\ref{tab:snlist}) or with the aim of studying stellar variability \citep{Conroy2018}. In this work, we attempt to obtain a distance to M~51 using two independent techniques; based on Cepheid variables (which is the first dedicated Cepheid study for M~51) and by applying an augmented version of EPM on SN~2005cs.

\begin{table}[]
    \centering
    \begin{tabular}{c c c}
        Supernova & Type \\
        \hline
        1994I & Ic\\
        2005cs & IIP (low-velocity)\\
        2011dh & IIb\\
        \hline
    \end{tabular}
    \caption{List of the supernovae observed so far in M~51. These were also followed up once or multiple times with \emph{HST}.}
    \label{tab:snlist}
\end{table}

Cepheids are well-known pulsating variable stars providing one of the most robust and simple distance estimation methods through the Leavitt period-luminosity relation \citep{Leavitt1912}, which made these stars the backbone of extragalactic distance measurements and Hubble constant estimations (e.g. \citealt{Riess2019, Riess2022}, see Sect.~\ref{sec:cep_method}). On the other hand, Type II supernovae also provide a well-established way to determine distances through the EPM \citep{Kirshner1974}. This technique provides an independent distance estimation that can be used for galaxies even in the Hubble flow (see Sect.~\ref{sec:EPM}). As opposed to the Cepheid method, which was not applied to M~51 previously, the EPM has been used multiple times in the literature on SN~2005cs to constrain the distance of this galaxy \citep{Takats2006, Dessart2008, Vinko2012, Bose2014}. However, as we describe in Sect.~\ref{sec:EPM} further, the method has undergone several improvements recently, which increased its accuracy \citep{Vogl2020}. Hence, pairing up and comparing the results of this augmented EPM with the first-ever Cepheid-based measurement for M~51 offers an excellent way to narrow down the distance to this galaxy and provides two independent distance estimates that can serve to understand the respective systematic uncertainties.

This paper is structured as follows: in Sect.~\ref{sec:data} we review the data we adopted for our analysis, in Sect.~\ref{sec:methods} we introduce and provide a brief outline for the different techniques, in Sect.~\ref{sec:cepheids} and Sect.~\ref{sec:2005cs} we present the individual steps of the analysis for the Cepheids and SN 2005cs, then in Sect.~\ref{sec:discussion} we discuss these results and summarize our findings.

\section{Data}
\label{sec:data}

To obtain a Cepheid distance to M~51, we made use of the catalogue and data presented in \cite{Conroy2018} (hereafter C18), which were derived based on 34 epochs taken between October 2016 and September 2017 by the \emph{Hubble Space Telescope} (\emph{HST}) during Cycle 24, using the Advanced Camera for Surveys ($ACS$). The photometric data published in C18 were derived using the DOLPHOT software package \citep{Dolphin2000}. Although this is a different software from the one that is regularly used for the reduction in Cepheid distance studies (namely DAOPHOT, e.g. \citealt{Riess2022}), the steps taken for the brightness and error budget estimation (such as the PSF model estimation or the artificial star tests) match the procedure described in the latest Cepheid works (e.g. \citealt{Yuan2022}). The applied photometric techniques were also tested before in the context of the \emph{HST} PHAT survey \citep{Dalcanton2012}, showcasing its high quality. A recent work has also investigated the consistency of the different reduction methods (DAOPHOT and DOLPHOT), finding good agreement \citep{Jang2023}. To ensure the precision of the reduction, C18 adopted the parameter settings from \cite{Dalcanton2012} (such as the aperture size, the detection threshold, and the maximal step size for positional iteration, among others; for the complete list, see Table~4 in \citealt{Dalcanton2012}).

The final C18 catalogue consists of F606W and F814W observations at up to 34 epochs for $\sim 72000$ stars. Apart from presenting the photometry, C18 carried out a Lomb-Scargle method-based analysis for the observations \citep{Lomb1976, Scargle1982} with the goal of studying long-term stellar variability in M~51. This analysis showed the presence of a period-luminosity relation, indicating that numerous Cepheids were also observed and that the data can be utilized for the determination of the distance to M~51. We emphasize that the number of available epochs in this catalogue is significantly higher than what is usually available for extragalactic Cepheids, making C18 an exceptionally rich dataset.

To derive a supernova distance we applied an augmented version of EPM to the data of SN~2005cs. SN~2005cs is a Type IIP supernova (SN IIP), which was discovered on 2005 June 28 by \cite{Kloehr2005}. It was followed thoroughly both photometrically and spectroscopically (by e.g. \citealt{Pastorello2006, Pastorello2009}). We adopted the photometric data obtained by the Katzman Automatic Imaging Telescope (KAIT, \citealt{Filippenko2001}) owing to its good coverage and quality, along with early photometric observations form amateur astronomers as collected by \cite{Pastorello2009}. We included spectroscopic observations from multiple sources that were taken in the epoch range required for our EPM approach, as described in Sect.~\ref{sec:EPM}: one spectrum from the Shane telescope\footnote{Shane = 3m Donald Shane Telescope, Lick Observatory, California (US)} using KAST, additional three from the Ekar telescope\footnote{Ekar = 1.82-m Copernico Telescope, INAF – Osservatorio di Asiago, Mt Ekar, Asiago (Italy)} obtained using the AFOSC spectrograph, one from the Swift satellite\footnote{Neil Gehrels Swift Observatory, NASA}, one obtained by TNG\footnote{TNG = 3.5-m Telescopio Nazionale Galileo, Fundacións Galileo Galilei – INAF, Fundación Canaria, La Palma (Canary Islands, Spain)} DOLORES, and finally one obtained by the P200 telescope\footnote{P200 = Palomar 200-inch Hale Telescope, Palomar Observatory - Caltech, Palomar Mountain, California (US)} using DBSP. For more details on the data, we refer the reader to \cite{Pastorello2009}.

\section{Methods}

In this section, we discuss the methods used to infer the distance to M~51. It is important to note that the presented methods are completely independent of one another; in particular, the supernova-based method requires no input or calibration from other techniques and acts as a primary distance estimator.

\label{sec:methods}
\subsection{Cepheid Period-Luminosity Relation}
\label{sec:cep_method}
The Cepheid period-luminosity relation (PL relation hereafter, \citealt{Leavitt1912}) is the well-known correlation between the pulsation period and the luminosity of Cepheids that has already been used for extragalactic distance determination more than a century ago \citep{Hertzprung1913}, and which provide the backbone of precise distance estimations today \citep{Riess2022}. The physical background of this relation and its general usage has been described in detail multiple times in the literature (see e.g. \citealt{Sandage1968, Madore1991, Freedman2001, Hoffmann2016}). Although easily applicable, as one only requires to measure the brightness and pulsation period of Cepheids, a significant limitation of the method is the effect of reddening, which increases the scatter in the PL diagrams. To remedy this, observations are usually taken at least in two separate filters, which are then used to calculate reddening-free magnitudes through the so-called Wesenheit function \citep{Madore1982}:

\begin{equation}
    \label{eq:Wesenheit}
    W_{VI} = m_{I} - R_{I,V-I}(m_V - m_I)
\end{equation}

\noindent where $m_V$ and $m_I$ denote the $V-$ and $I-$band magnitudes, while $R_{I,V-I} = A_I/E(V-I)$ denotes the total-to-selective extinction ratio. The Wesenheit function provides the greatest increase in quality when near-infrared observations are available, which are intrinsically less sensitive to the reddening, and make the Wesenheit magnitudes more robust against an incorrectly assumed reddening law. Nevertheless, even the optical two-band Wesenheit indices reduce the effect of reddening significantly. For our work, we employed two Wesenheit functions for the specific \emph{HST} $F555W$, $F606W$ and $F814W$ bands:

\begin{equation}
\label{eq:W555-814}
    W_{F555W, F814W} = F814W - 1.261 \cdot (F555W - F814W)
\end{equation}
\noindent and
\begin{equation}
\label{eq:W606-814}
    W_{F606W, F814W} = F814W - 1.757 \cdot (F606W - F814W)
\end{equation}

\noindent where the total-to-selective extinction ratios were calculated for $R_V = 3.3$ assuming a \cite{Fitzpatrick1999} reddening law and a typical Cepheid SED following \citet{Anderson2022}.

For our analysis, we chose NGC~4258 as the anchor galaxy, whose distance is known to a high precision on a geometric basis (based on a maser, \citealt{Reid2019}) and which hosts several Cepheids \citep{Yuan2022}. The choice of NGC~4258 was also motivated by the fact that it was observed under similar conditions as M~51: the galaxies are at approximately the same distance, were observed by $HST$, with $ACS$, although using slightly mismatching filters. Hence their direct comparison allows for a differential distance estimation. Since NGC~4258 was observed in ($F555W,F814W$), but not in $F606W$ (which was available for M~51 instead of F555W), we only had to characterize the PL-relation in $W_{F555W,F814W}$ (after the relevant magnitudes were converted for the M~51 sample, see Sec.~\ref{sec:conversion}). Consequently, the observed M~51 F606W magnitudes and the corresponding Wesenheit function were only used directly for sample selection, as described in Sec.~\ref{sec:cepheids}.

For the distance estimation, following \cite{Riess2022}, the PL relation for M~51 can be written as

\begin{equation}
\label{eq:PL}
    [W_{F555W, F814W}]_i = \alpha \cdot (\log P_i - 1) + \beta + \mu_{0,M51} + \gamma \cdot [\mathrm{O/H}]_i
\end{equation}

\noindent where $i$ indexes the individual Cepheids in the sample, $W_{F555W, F814W}$ denotes the Wesenheit magnitude, $[\mathrm{O/H}]_i$ the metallicity with the usual $[\mathrm{O/H}] = 12 + \log(\mathrm{O/H})$ definition, $P$ the period of the Cepheids, and the parameters $\alpha$, $\beta$ and $\gamma$ define the empirical relationship. The parameters $\alpha$ and $\beta$ can be determined by fitting the PL relation of the anchor galaxy, NGC~4258, by adopting its distance modulus ($\mu_{0,NGC4258} = 29.397 \pm 0.032$ mag, \citealt{Reid2019}). For  $\gamma$ we adopt the  value of $\gamma = -0.201 \pm 0.071$ mag/dex corresponding to the (F555W,F814W) filter set from \cite{Breuval2022}. This value is slightly different from, but also consistent with the factor used by \cite{Riess2022}, $Z_W = -0.251 \pm 0.05$, which is applicable for the $W_H$ Wesenheit magnitudes. However, given the 0.4 dex average metallicity difference between M~51 and NGC~4258 \citep{Zaritsky94,Yuan2022}, such a difference between $Z_W$ and $\gamma$ would only lead to a very small distance offset of 1\%, hence the exact choice of metallicity factor has no major influence on the final estimate.

\subsection{Tailored-EPM}
\label{sec:EPM}
To estimate the distance to M~51 based on SN~2005cs, we applied a variant of the tailored expanding photosphere method (tailored-EPM hereafter, \citealt{Dessart2006, Dessart2008, Vogl2020}). The method itself is an augmented version of the classic EPM, which is a geometric technique relating the photospheric radius of the supernova to its angular diameter \citep{Kirshner1974}. Although straightforward, the classic method is prone to several systematics and uncertainties \citep{Jones2009}. As pointed out by \cite{Dessart2005}, these uncertainties can only be reliably suppressed if one estimates the relevant physical parameters for the EPM analysis through the complete radiative transfer-based modelling of the supernova spectra (which is referred to as tailoring the EPM estimation). We call this augmented version the tailored-EPM analysis, which bears many similarities to the Spectral-fitting Expanding Atmosphere Method (SEAM) introduced by \cite{Baron2004}.

For the required spectral modelling of the supernova we make use of the spectral emulator developed by \cite{Vogl2020}, which is based on radiative transfer models calculated with a modified and Type II supernova-specific version of TARDIS \citep{TARDIS, Vogl2019}. The emulator not only reduces the time required for the spectral fitting significantly but also yields precise estimations of physical parameters on a maximum likelihood fitting basis. The background of this spectral fitting method has been thoroughly described in \cite{Vogl2020}, while its application and the required calibration steps were summarized in \cite{Csornyei2022}. It has already been showcased in \cite{Vogl2020} and \cite{Vasylyev2022}, and has also been shown to provide internally consistent results for sibling supernovae \citep{Csornyei2022}. For our work, we reapplied the steps detailed in these articles, although for completeness we summarise them in Sect.~\ref{sec:2005cs}.

For the distance measurement, one has to first estimate the photospheric angular diameter of the supernova ($\Theta = R_{\textrm{ph}}/ D$, where $R_{\textrm{ph}}$ denotes the radius of the photosphere and $D$ the distance measured in Mpc) for each spectral epoch. This estimation is done by minimizing the difference between measured and model apparent magnitudes ($m^{\textrm{obs}}$ and $m$, respectively) at the given epoch using $\Theta$ as argument:

\begin{equation}
\label{eq:EPM_mod}
    \Theta^{*} = \arg \min_{\hspace*{-15pt}\Theta} \sum_{\textrm{S}}\left(m_{\textrm{S}} - m_{\textrm{S}}^{\textrm{obs}}\right)^2
\end{equation}

\noindent for all available photometric bands $S$. To estimate the model apparent magnitudes for each of these bands, one has to employ the distance modulus formula and replace the distance with the angular diameter as follows:

\begin{equation}
\label{eq:EPM}
\begin{split}
    &m_S - M_S = - 5 + 5 \log (D) + A_S\\
    &m_S = M_S - 5 + 5 \log \frac{R_{\textrm{ph}}(\Sigma^{*})}{\Theta(\Sigma^{*})} + A_S\\
    &m_S = M_S^{\textrm{ph}}(\Sigma^{*}) - 5 \log [\Theta(\Sigma^{*})] + A_S \\
    & \textrm{with}\\
    &M_S^{\textrm{ph}}(\Sigma^{*}) = M_S + 5 \log \frac{R_{\textrm{ph}}(\Sigma^{*})}{10\textrm{ pc}}
\end{split}
\end{equation}

\noindent Here, $\Sigma^{*}$ denotes the set of physical parameters corresponding to the best fit, $M_S^{\textrm{ph}}$ is the absolute magnitude predicted by the radiative transfer model at the position of the photosphere and $A_S$ denotes the broadband dust extinction in the bandpass. It is important to note, that the distance $D$ is the only free parameter, which is not directly determined by the spectral fits. With this definition, the best-fitting angular diameter $\Theta^{*}$ can be determined for each of the relevant spectral epochs.

Finally, assuming the ejecta is in homologous expansion ($R_{\textrm{ph}} = v_{\textrm{ph}}t$), the distance to the supernova and its time of explosion can be estimated through a Bayesian linear fit to the ratios of the angular diameters and the photospheric velocities ($\Theta / v_{\textrm{ph}}$) versus time $t$. In the fit, we assume Gaussian uncertainties for $\Theta / v_{\textrm{ph}}$ of 10\% of the measured values for a given colour excess following \cite{Dessart2006}, \cite{Dessart2008}, and \cite{Vogl2020}. We set a flat prior for the distance, whereas for the time of explosion we use the normalized histogram of the $t_0$ posterior from a fit to the early light curve as the prior (following \citealt{Csornyei2022}). However, instead of applying the standard $\chi^2$ based likelihood for the EPM, we used the modified fitting approach from \cite{Csornyei2022} to take into account the correlated errors caused by the reddening. Essentially, with this approach, we evaluate the EPM on multiple reddening values, drawn from the distribution of the single epoch best-fit $E(B-V)$ values (see Sect.~\ref{sec:specmod} on how these values are obtained). This approach in the end yields a more realistic uncertainty on the EPM distance.


\section{Cepheids}
\label{sec:cepheids}
In order to obtain a proper understanding of the data presented in C18, and to ensure the good quality of the Cepheid sample, we chose to first reanalyse the catalogue starting from the light curves. This, in turn, allowed us to filter the dataset in multiple steps, instead of applying cuts in the colour-magnitude diagram alone.

\subsection{Filtering and sample selection}
\subsubsection{Period filtering}
\begin{figure}
    \centering
    \includegraphics[width = \linewidth]{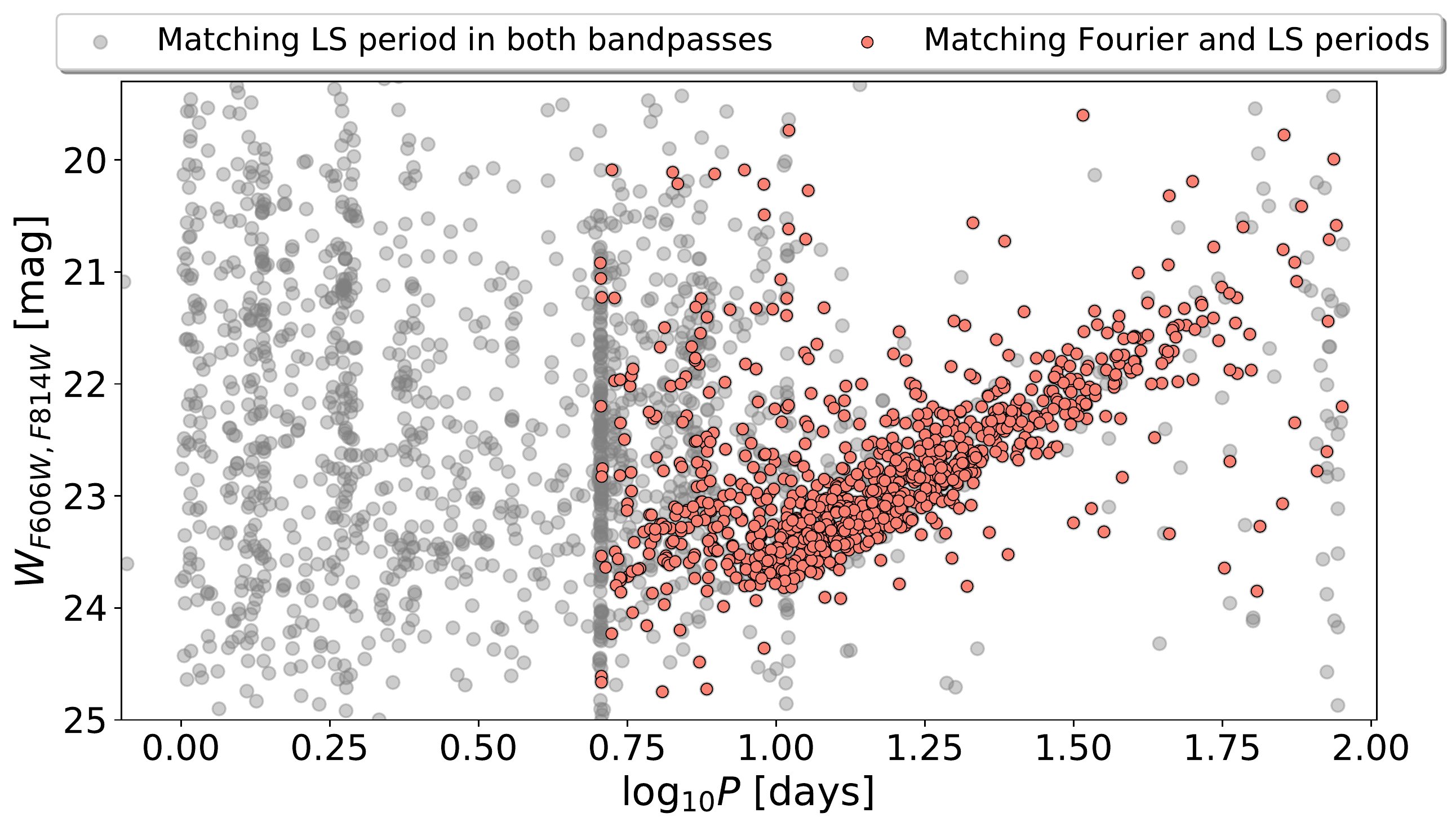}
    \caption{Catalogue PL relation after the filtering based on the Lomb-Scargle (LS) and Fourier periods and the Nyquist frequency. The Wesenheit indices were calculated based on the catalogue magnitudes. The grey dots show the stars where the LS period matched between the F606W and F814W bands, whereas the red dots show the stars for which the LS and Fourier periods matched for all filters.}
    \label{fig:period_filt}
\end{figure}

As a first quality estimation for the sample stars, we inspected the robustness of their period, if any, for the entire C18 catalogue. For this step, we compared the periods from C18 obtained by the Lomb-Scargle method in the $F606W$ and $F814W$ filters with one another, and with the equivalent values we obtained using Discrete Fourier Transformation (DFT, \citealt{Deeming1975}) employing \texttt{Period04} \citep{Lenz2005}. Given that the period should be independent of the chosen bandpass or analysis method, it can be used for an initial quality estimation. If any of the four obtained values deviated by more than $1\%$ from the rest, we removed the corresponding star from the sample. Furthermore, additional outliers were removed based on the length of the calculated periods. Even though the C18 sample is extremely rich compared to other extragalactic light curve samples, it is still reasonably sparse; on average, one datapoint was taken every 10 days. This puts a limit on the maximal non-aliased frequency that can be estimated using C18 \citep{Nyquist1928,Eyer1999}. We thus calculated the Nyquist period for each of the remaining sources ($P_{Nyq.} \approx 5-10$ days) and kept only those, where the estimated period was longer. This single step reduced the sample size from $\sim$72000 to merely 950. Fig.~\ref{fig:period_filt} shows the period-Wesenheit plot for the sample after the period filtering. The plotted Wesenheit values were calculated based on Eq.~\ref{eq:W606-814}, using the C18 catalogue magnitudes.

\subsubsection{Light curve shape filtering}
After removing non-periodic stars from the catalogue, we attempted to limit our sample further based on their light curves to make sure that only Cepheids or variables with sinusoidal light curves are carried forward. The main reason for such a filtering is that light curves give a more reliable handle on the variable's type than the colour and the brightness; apart from not being sensitive to reddening, this filtering also removes non-Cepheid stars that seem to scatter into the instability strip region due to reddening or blending. Given that most of the sample stars were observed at 34 epochs, which is significantly more than what is available for Cepheids in other galaxies, we could perform a more detailed light curve shape analysis. Instead of inspecting the light curves visually, and selecting the Cepheid stars by hand, we applied an automated and reproducible light curve filtering technique.

\begin{figure*}
    \centering
    \includegraphics[width = \linewidth]{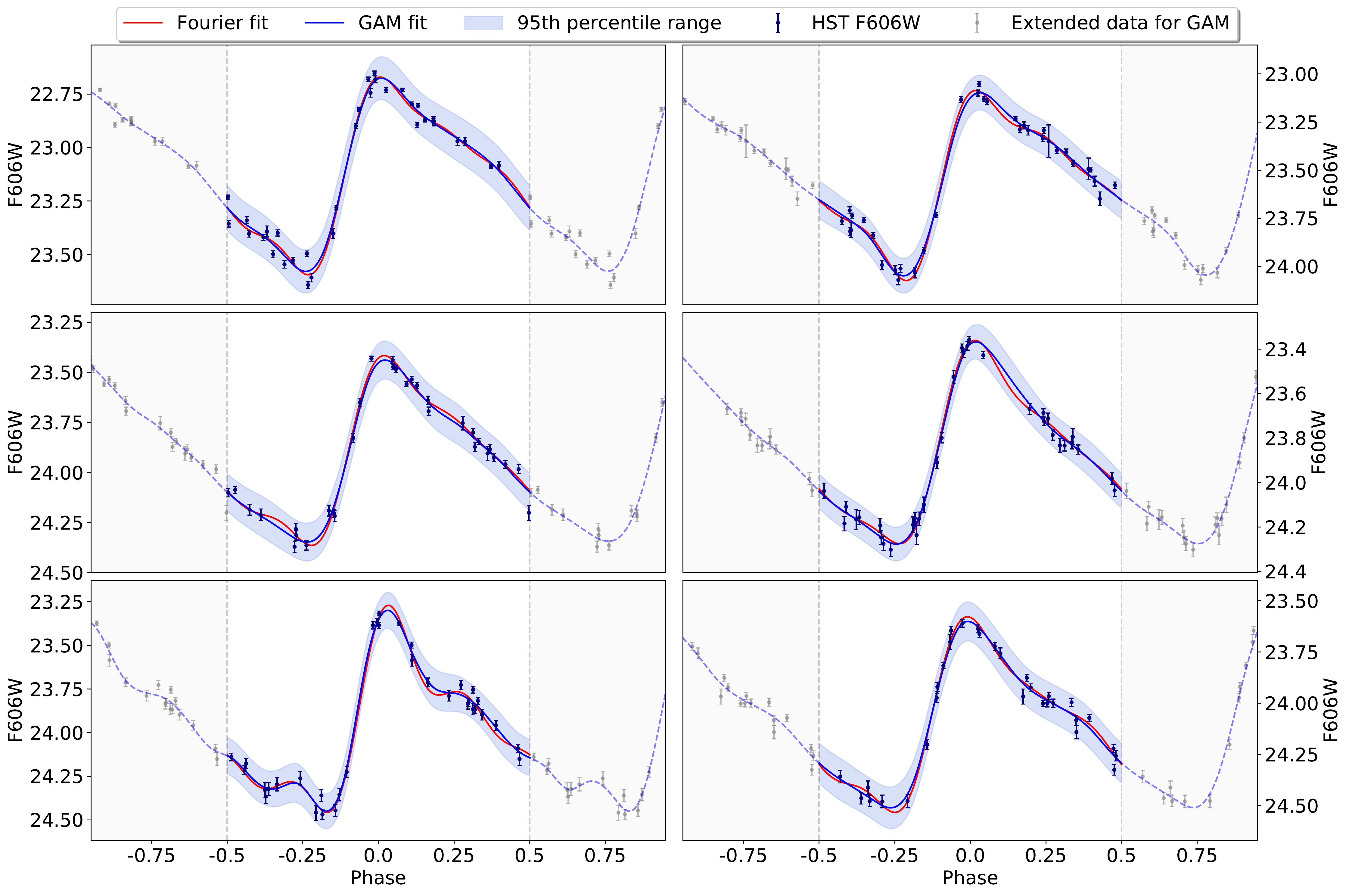}
    \caption{Example of GAM and Fourier fits for a handful M~51 Cepheids. The blue-shaded region shows the 95th percentile range of the GAM fits; as seen on the plots, the two models agree perfectly within these uncertainty limits. The grey shaded region shows the extension region for the GAM fitting, which was used to enhance the stability of the phase curve fits towards the edges of the region of interest (close to the phases -0.5 and 0.5).}
    \label{fig:exCep}
\end{figure*}

We chose to build this step on Principal Component Analysis (PCA, \citealt{Pearson1901}). PCA is a commonly used tool for reducing the dimensionality of data, hence allowing for a fast and cheap classification and comparison of dataset elements (see, e.g., \citealt{Dobos2012,Bhardwaj2016,Seli2022}). As a reference sample, we adopted the Cepheid set from \cite{Yoachim2009}, where they applied PCA to obtain a reliable template set of Cepheid light curves. This set consists of Milky Way and LMC $V$ and $I$ band light curves which were covered well enough in phase to allow for a detailed light-curve shape analysis. Since the \cite{Yoachim2009} sample of Cepheids was measured in the Bessell system, we converted the M~51 measurements from the $ACS$ system before the light curve fitting based on the relations described in \cite{Sirianni2005}.

To apply PCA, we first fit the light curves, both in the reference \cite{Yoachim2009} and the M~51 C18 sample. For this purpose, we applied Generalized Additive Models (GAMs, \citealt{GAM}). GAMs are smooth semi-parametric models, in the form of a sum of penalized B-splines, which allow modeling non-linear relationships with suitable flexibility. The method itself bears similarities with Gaussian Process fitting, technically using splines as the kernel function. This allows for a non-parametric, smooth, yet robust fitting method that takes into account the data uncertainties. For the implementation of this method, we made use of the Python package \texttt{pyGAM} \citep{pygam}. While the Fourier method would also perform perfectly for Cepheid light curves, it may not handle other types of variables, such as eclipsing binaries, with similar precision. Furthermore, the number of the Fourier components has to be varied from star to star, in order to avoid overfitting. This, however, can lead to biases when all the light curve models of different complexities are inspected together. Here GAM models provide a viable alternative, as they yield smooth light curve models whose complexity is automatically set by the data quality. This property makes the technique favourable for generic light curves.

To fit the light curves using GAMs, we first calculated their phase curves using the C18 periods. A comparison of the GAM and DFT fitting can be seen in Fig.~\ref{fig:exCep} for a selected few M~51 Cepheids. The two models agree well for good quality time series. In worse cases, however, in the presence of outliers or larger uncertainties, the naive Fourier method performs worse and increases the model complexity (i.e., overfits the data), as shown in Fig.~\ref{fig:app_comp}. However, the GAM model remains smooth and avoids overfitting, hence it allows for reliable filterings in more uncertain cases as well.

\begin{figure}
    \centering
    \includegraphics[width = \linewidth]{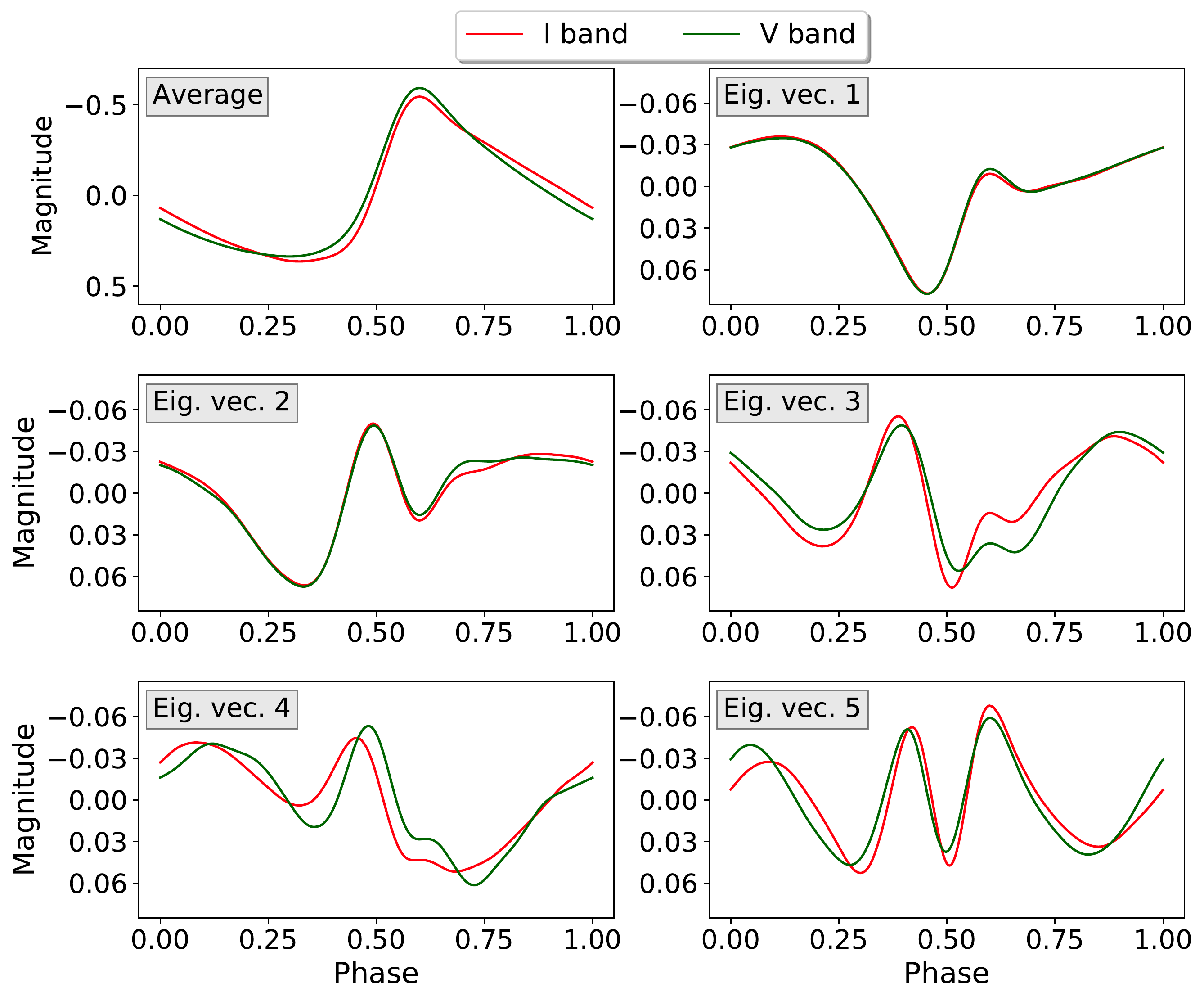}
    \caption{Basis vectors obtained from the PCA applied to the model light curves of the reference sample. The green curves show the Bessell V-band vectors, and the red curves the I-band vectors.}
    \label{fig:PCA}
\end{figure}

This light curve fitting was carried out both for the reference and M~51 sample stars. After this step, each of the light curves was rephased, so that their maximal brightness would fall on the same phase value. We then performed the PCA on the reference sample, obtaining the average curve and the eigenvectors required later on. These vectors are shown in Fig.~\ref{fig:PCA}. For our analysis, we made use of only the first five eigenvectors obtained by the PCA, as they contained $>$95\% of the variance. This is similar to the choice made by \cite{Yoachim2009}, who retained only the first four principal components (although they only aimed to explain $\geq$90\% of the variation).

 The basis was then used to expand the M~51 model phase curves (after applying the same phase normalization), obtaining their expansion coefficients. These expansion coefficients were then used to filter out non-Cepheid variables of the sample. For this, we set up a grid in the field of expansion coefficients and counted the number of stars that fell in the individual grid elements (i.e. we set up a multi-dimensional histogram for the coefficients). Then, we removed every star for which the expansion coefficients fell in bins not covered by the reference sample stars. This approach is similar to defining a Convex Hull for the reference sample and using that for the removal of unfit elements, with a lower resolution. This step ensured that only Cepheid-like or sinusoidal light curves remained in the sample.

\subsubsection{Instability strip filtering}
\label{sec:isf}
As a final step of filtering, we removed stars from the sample that exhibited colours that were too blue for Cepheids or were significantly redder than the instability strip. The position of the bluer stars on the colour-magnitude diagram (CMD) cannot be explained by extinction; hence, these stars are either non-Cepheid variables or Cepheids strongly blended with clusters \citep{AndersonRiess2018}. On the other hand, the stars on the redder end can be significantly reddened Cepheids. While the use of the Wesenheit system solves most of the errors tied to the reddening, removing such highly reddened stars will reduce our exposure to the limitations of the Wesenheit relation, resulting in a smaller scatter in the PL relation. To this end, we have constructed the cumulative density function (CDF) of the CMD, by moving a dividing line along the colour axis. The slope of this dividing line was adopted from \cite{Riess2019}, to make sure it matches the instability strip. 

\begin{figure}
    \centering
    \includegraphics[width = \linewidth]{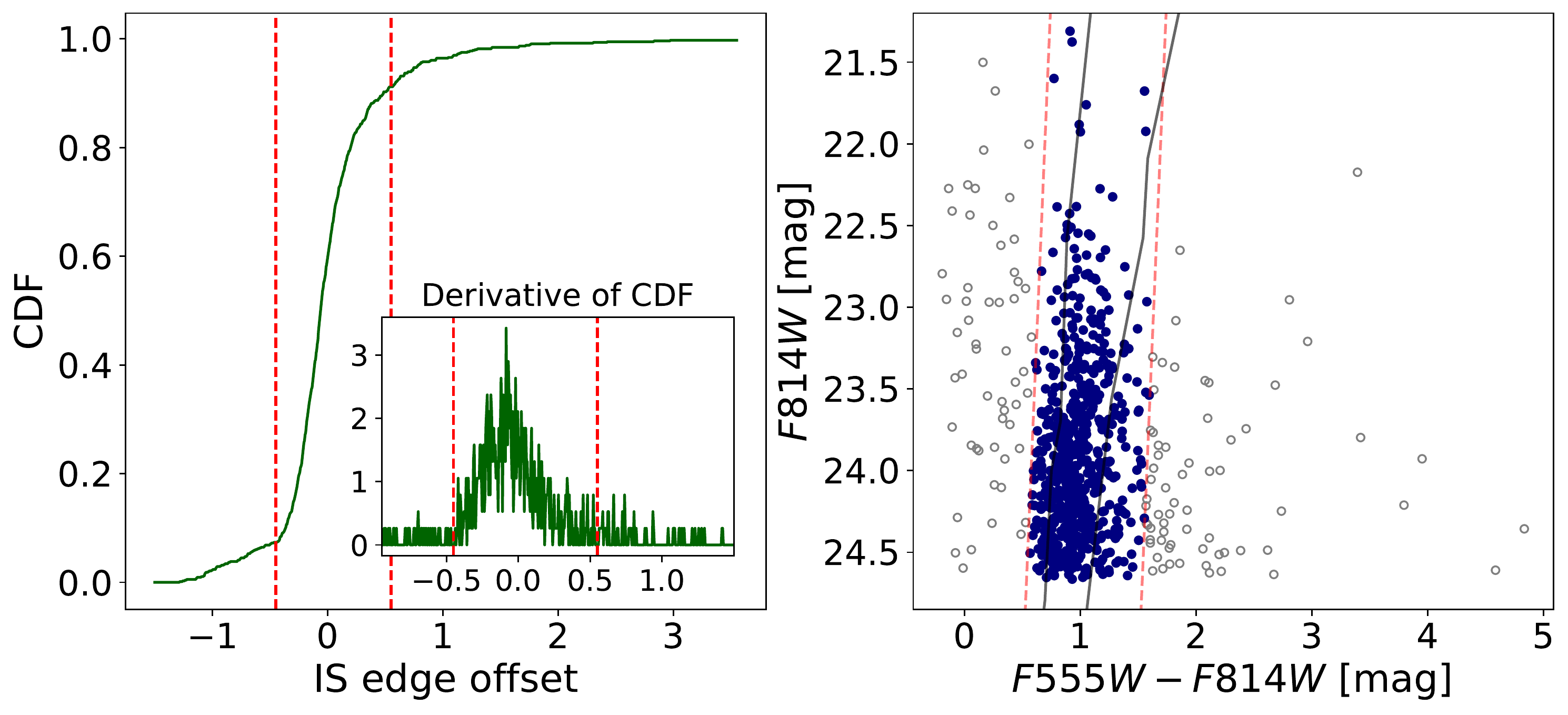}
    \caption{(\textbf{Left:}) CDF of the CMD parametrized by the offset of the adopted instability strip (IS) slope. The inset shows the derivative of the CDF in the range of interest. The red dashed lines show the limits where the CDF values start to either increase significantly or level off, i.e. the positions of the edges of the M~51 IS. (\textbf{Right:}) The CMD of the M~51 sample, with the IS edge derived based on the CDF (red line). The grey points show the stars which were flagged as outliers. The grey curves show the theoretical instability strip edges from \cite{Anderson2016} reddened by the Galactic colour excess of 0.03 mag for comparison.}
    \label{fig:cmd}
\end{figure}

To derive the CDF, we counted the stars left from this dividing line on the CMD, while moving it from bluer colours towards the redder ones. For the bluer edge of the instability strip, we assumed that there is a given offset value for this line, at which all stars on the left of it are outliers, while the ones on its right are likely Cepheids (see the right plot of Fig.~\ref{fig:cmd}). To find this right offset value, we calculated the first derivative of the CDF. We expect this derivative to initially be flat, as long as the dividing line is to the left of the instability strip on the CMD; in this case, at each step, it only moves over a few stars. However, when it enters the instability strip, the number of stars moved over at each step would drastically increase; this shows up in the derivative curve as an upturn. Hence, the best position for the dividing line can be set by this upturn; choosing the offset value corresponding to it will ensure that the filter does not enter the instability strip, thus not cutting bona-fide Cepheids away, while removing as many outliers as possible. A similar scenario was followed on the red edge of the instability strip; however, here, instead of looking for the rise in the CDF derivative, we attempted to find the offset where it levelled off. This procedure is shown in Fig~\ref{fig:cmd}, with the red lines marking the filter. The optimal offset value was found to be -0.45 for the blue edge, and 0.55 for the red edge in this setup. As seen on the plot, the resulting Cepheid set aligns well with the reddened theoretical instability strip edges of \cite{Anderson2016} (with galactic reddening applied), while cutting away the outliers.

\begin{figure}
    \centering
    \includegraphics[width = \linewidth]{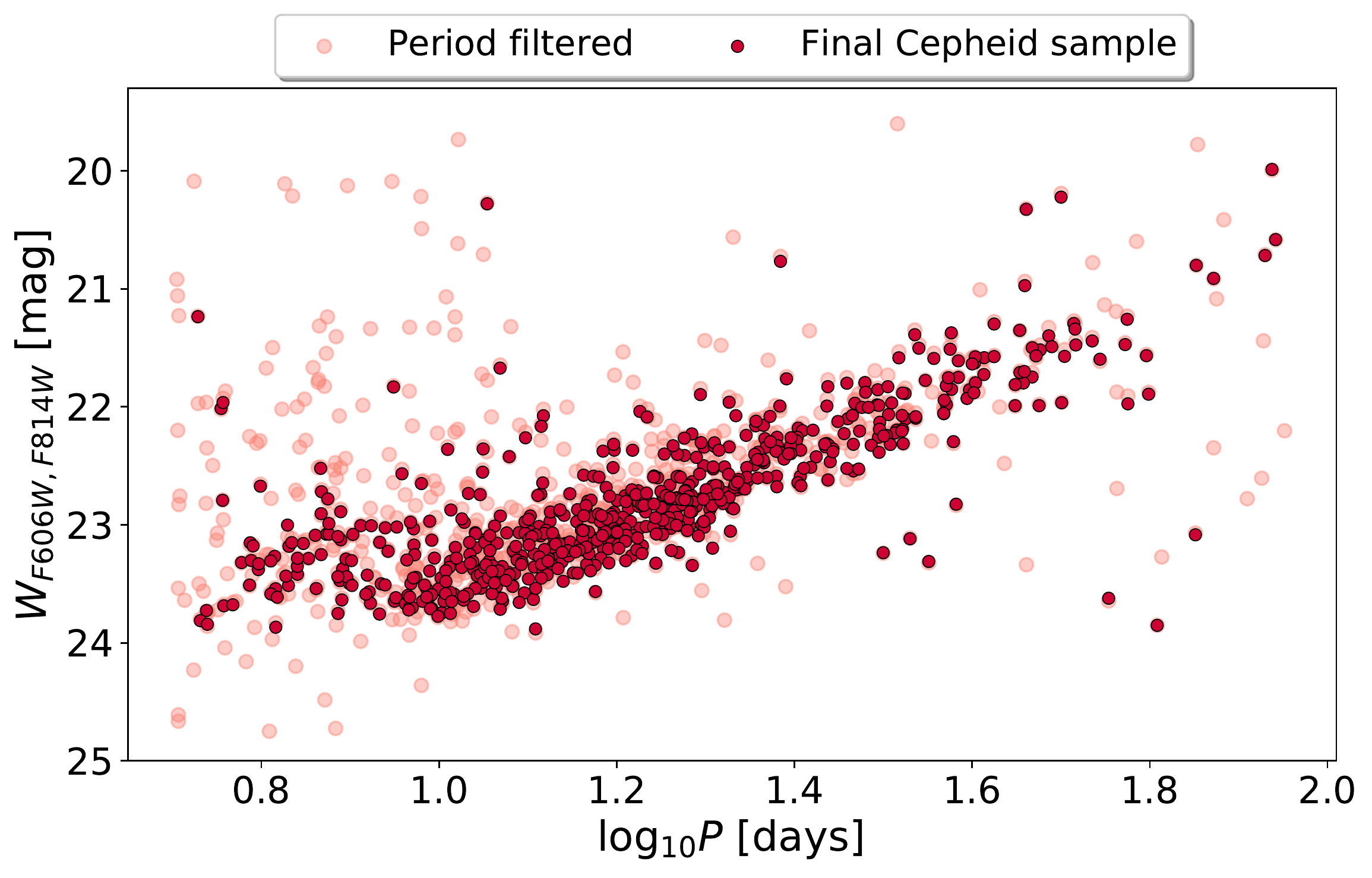}
    \caption{Final set of M~51 Cepheids and their period-luminosity relation. The grey dots show the resulting sample after the first filtering step, while the red points correspond to the final M51 Cepheid set after the light curve shape and the CMD-based selections.}
    \label{fig:PL-fin}
\end{figure}

The three filters yielded a sample of M~51 variables that are most likely Cepheids, in an almost completely automated manner. In total, the entire catalogue of variable stars was narrowed down from $\sim 72000$ stars from all types to 638 Cepheids. Table~\ref{tab:steps} shows how the number of sample stars changed after the individual filters. As a final step, we used the GAM models to re-calculate the flux-averaged magnitudes for each of the Cepheids. The magnitude uncertainties were also re-evaluated based on the GAM light-curve fit confidence intervals. The final sample of Cepheids and the corresponding period-luminosity plot can be seen in Fig.~\ref{fig:PL-fin}. As seen in the plot, numerous Cepheids are available for the distance determination, and although the majority of the stars are fundamental mode Cepheids, the overtone branch is also populated and well-distinguishable. The positions of the Cepheids in the final sample within M~51 are shown in Fig.~\ref{fig:m51-ceps}. The final list of M~51 Cepheids is given in Tab.~\ref{tab:appendix} in the Appendix.

\begin{table}[]
    \centering
    \begin{tabular}{c c}
    \hline
    Filtering step & \# of sample stars \\
    \hline
    - & 72623\\
    Period matching  & 950\\
    Light curve filtering & 759\\
    Instability strip cut & 638\\
    \hline
    \end{tabular}
    \caption{Changes in the Cepheid sample size after the individual filtering steps.}
    \label{tab:steps}
\end{table}

\begin{figure}
    \centering
    \includegraphics[width = \linewidth]{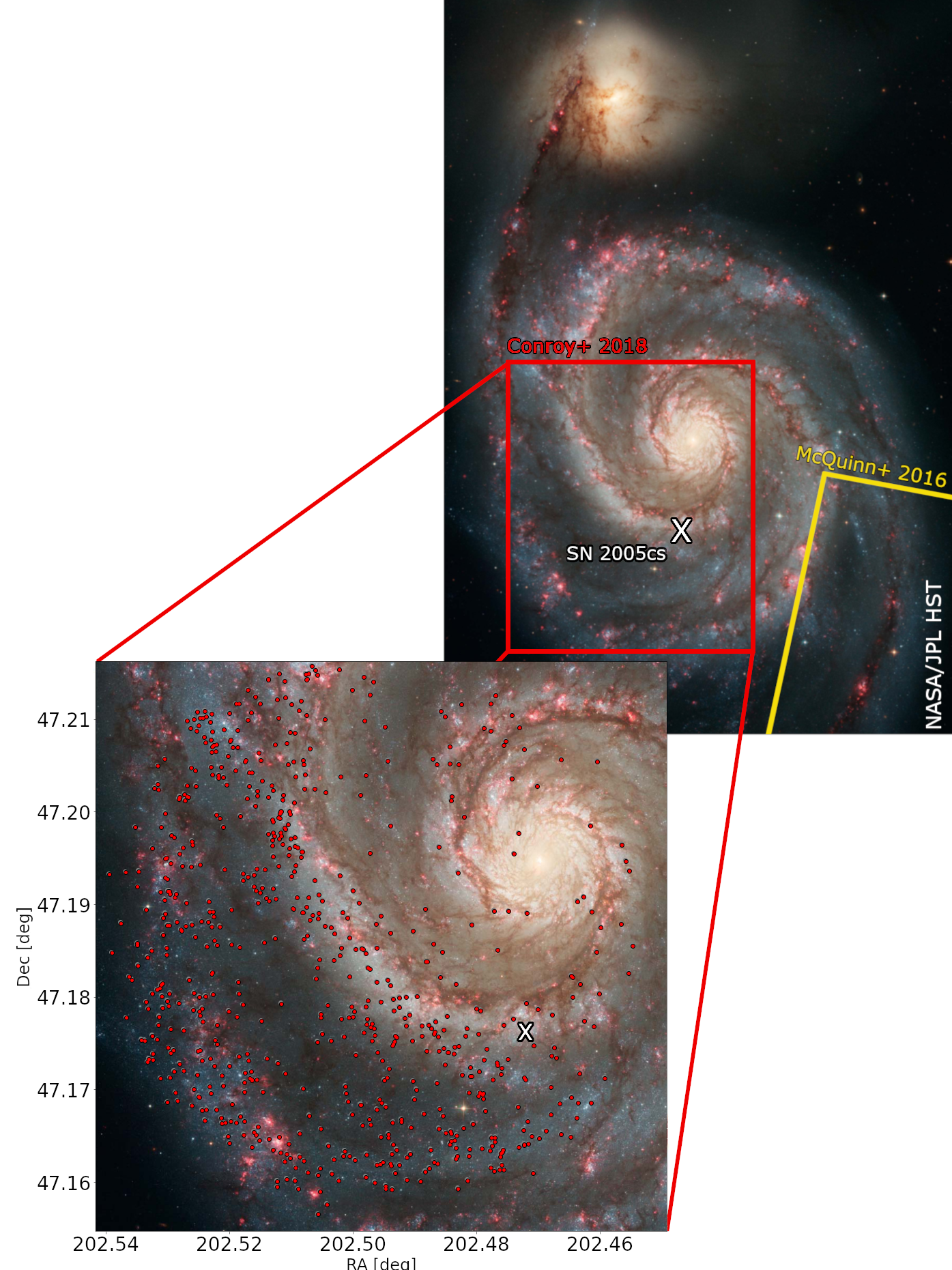}
    \caption{Positions of the final sample of Cepheids within M~51. The background image was taken by the \emph{Hubble Space Telescope}. The red box indicates the field observed by C18. The yellow shape shows the position and the orientation of the M~51 field used for the TRGB by \cite{McQuinn2016}. The white cross denotes the position of SN~2005cs.}
    \label{fig:m51-ceps}
\end{figure}

\subsection{$F606W$ - $F555W$ magnitude conversion}
\label{sec:conversion}
\begin{figure}
    \centering    
    \includegraphics[width = \linewidth]{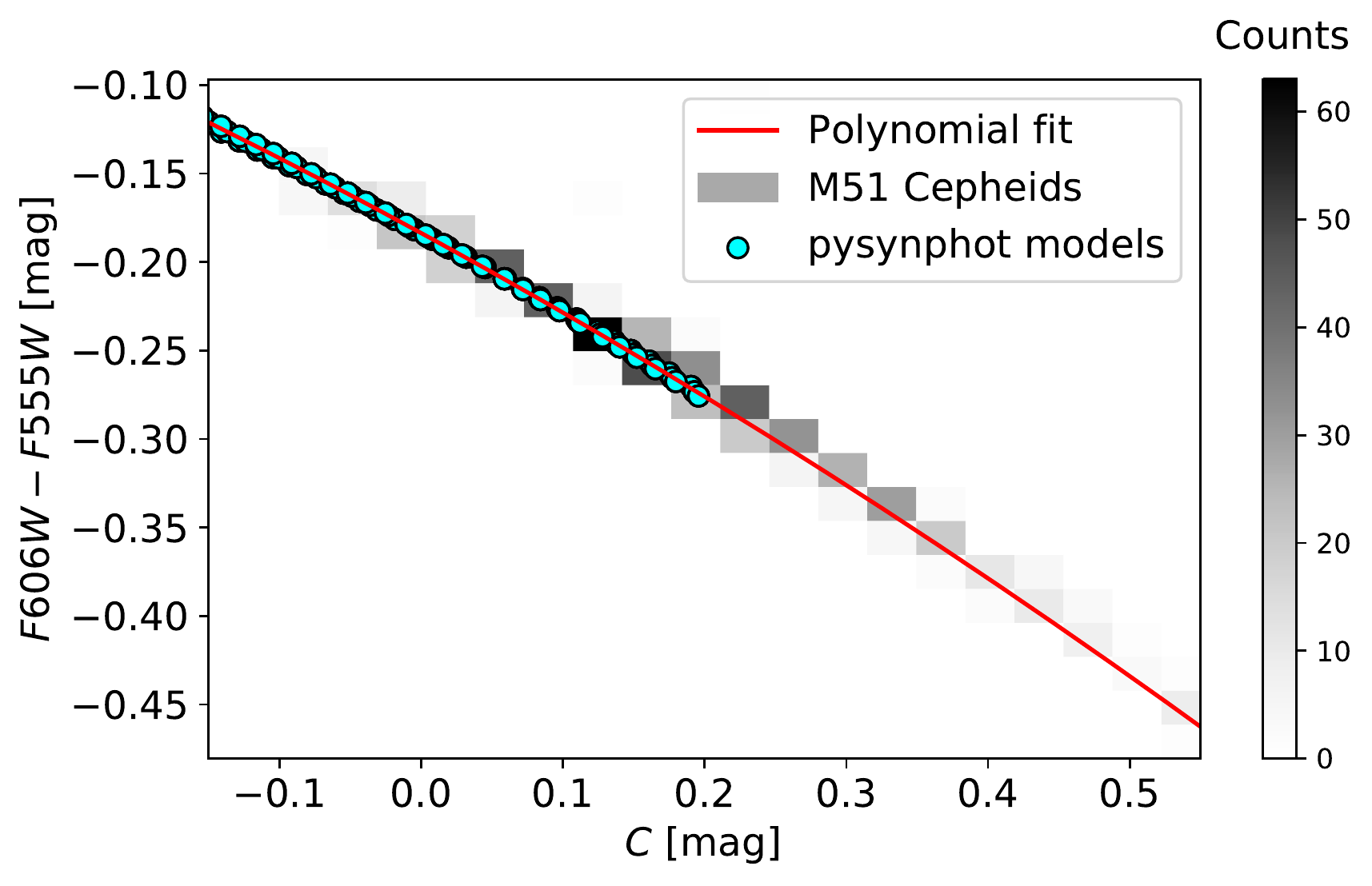}
    \caption{Synthetic \texttt{pysynphot} $HST$ colours and the fitted trend, which was subsequently used for transformation. The $C$ denotes the $F606W - F814W$ colour, offset by the sample average. The grey-shaded background shows the distribution of the M~51 Cepheid average colours after applying conversion and refitting the light curves. Given that our model grid only extended down until 5000 K in temperature, the synthetic data do not cover all colours seen in the catalogue (as some stars scatter up to redder colours due to the reddening). Nevertheless, even for the redder stars, we extrapolate an F555W magnitude using the given transformation curve.}
    \label{fig:pysynphot}
\end{figure}

To determine the distance modulus relative to NGC~4258, the $F606W$ observations for the M~51 sample required conversion to $F555W$. To carry this out, we made use of the spectral library of ATLAS9 stellar atmosphere models \citep{Castelli2003} accessible through \texttt{pysynphot} \citep{pysyn}. After limiting the models to the temperature range of Cepheids ($5000 \textrm{ K} < T_{\textrm{eff}} < 6500 \textrm{ K}$) we calculated the synthetic $F555W - F814W$ and $F606W - F814W$ colours. Then, by fitting the relation between the colours using a second-order polynomial, we inferred the $F555W$ brightnesses for the M~51 sample. We note that we carried out the conversion to each of the individual measurement epochs instead of just the Cepheid average magnitudes. The conversion curve and its fit are displayed in Fig.~\ref{fig:pysynphot}. The obtained transformation equation is

\begin{equation}
    F606W - F555W = -0.129 \cdot C^2 - 0.436 \cdot C - 0.183,
\end{equation}

\noindent where $C$ denotes the original colour, $F606W-F814W$, with the sample mean subtracted from it (to minimize the fitting and conversion uncertainties). From here on, we only used the converted $F555W$ along with the original $F814W$ values for the analysis. Throughout the conversion, no reddening corrections were applied, even though this is known to impose an additional small uncertainty (given that the models and the colour conversion assume an extinction-free scenario). On the one hand, constraining this uncertainty properly is hard, since it requires knowledge of the internal (within M~51) reddening on a Cepheid-to-Cepheid basis. On the other hand, the good match found between the sample and the slightly reddened theoretical instability strip boundaries (Fig.~\ref{fig:cmd}) shows that the majority of the sample is not affected by strong extinction. Assuming a systematic reddening towards the M~51 Cepheids of $E(B-V) = 0.03$ mag (which is similar to the value found by our analysis of SN~2005cs, see Sect.~\ref{sec:2005cs}) would only result in a minor $F606W - F555W$ colour difference (smaller than $\sim 0.01$ mag), which is negligible compared to the other systematics affecting our results. Hence, we choose to neglect this term (especially, given that the highly reddened stars were already clipped using the instability strip boundary estimation presented in Sect.~\ref{sec:isf}).

\subsection{PL relation fitting}
\label{sec:PL}

\begin{figure}
    \centering
    \includegraphics[width = \linewidth]{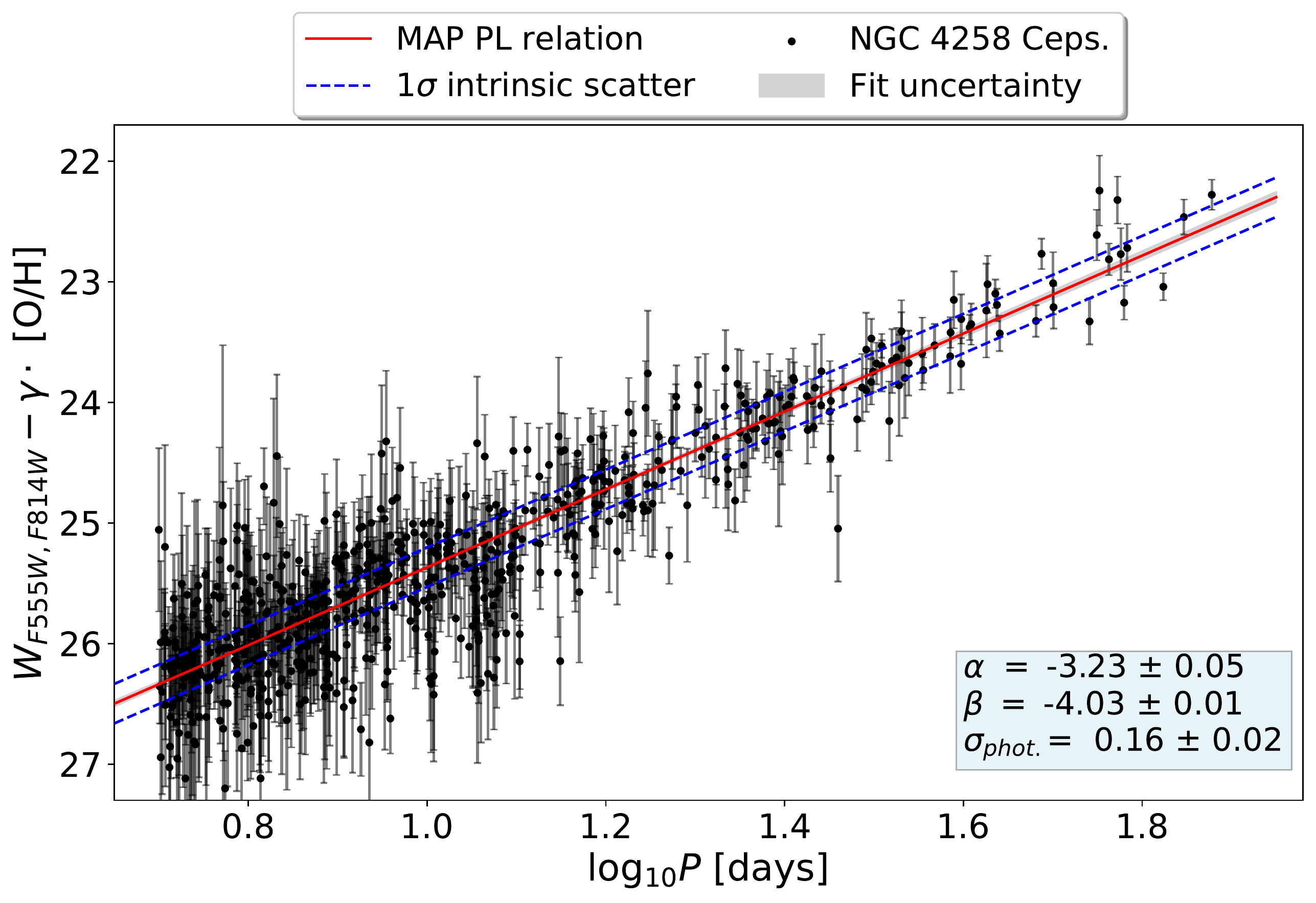}
    \caption{The PL relation fit to the NGC~4258 anchor data. The red line shows the maximum a posteriori (MAP) fit, while the dashed lines denote the range of intrinsic scatter, which is due to observational constraints and the finite width of the instability strip. For simplicity and to reduce scatter, we marginalized over the metallicity values for this plot.} 
    \label{fig:N4258}
\end{figure}

In order to measure the distance to M~51 based on the newly found Cepheids, we chose NGC~4258 as the anchor galaxy, which is the only maser host with observed Cepheid variables. We adopted a distance of $D = 7.576 \pm 0.082$ Mpc for NGC~4258 following \cite{Reid2019} (which corresponds to a distance modulus of $\mu = 29.397 \pm 0.032$ mag). The observed set of Cepheids in NGC~4258, their brightnesses in the HST bands and their metallicities were adopted from \cite{Yuan2022}. The data published in that paper was also obtained using $ACS$, hence no additional zero-point corrections were necessary. To estimate the reference PL relation fit parameters, which were later used for the distance estimation of M~51, we fit Eq.~\ref{eq:PL} to the NGC~4258 data set. To fit the model, we made use of the \texttt{UltraNest}\footnote{\url{https://johannesbuchner.github.io/UltraNest/}} package \citep{Buchner2021}, which allows for Bayesian inference on complex, arbitrarily defined likelihoods based on the nested sampling Monte Carlo algorithm MLFriends \citep{Buchner2016, Buchner2019}. This allowed us to modify the likelihood. As pointed out by for example \cite{Breuval2022}, due to the finite width of the instability strip, the Cepheids exhibit a non-negligible scatter around the "true" period-luminosity relation, which has to be included in the model. This intrinsic scatter is naturally further enhanced by the photometric uncertainties. To include it, we extended the $\chi^2$ likelihood according to \cite{Hogg2010} and applied it in the form

\begin{equation}
    \ln p(W_{VI} | \Omega) = - \frac{1}{2} \sum_i \left(\frac{W_{VI,i} - m_{\mathrm{model},i})^2}{\sigma ^2_i} + \ln{2 \pi \sigma^2_i} \right),
\end{equation}

\noindent where $W_{VI}$ denotes the observed $W_{F555W, F814W}$ Wesenheit magnitudes, $\Omega$ the set of fit parameters (slope $\alpha$ and offset $\beta$ of the linear fit), $m_{\mathrm{model}}$ are the model Wesenheit magnitudes as obtained from the PL relation, and $\sigma^2$ is the extended uncertainty, defined as the quadrature sum of the measurement uncertainty and the photometric scatter $\sigma^2_i = \sigma^2_{\mathrm{measurement},i} + \sigma^2_{\mathrm{photometric}}$. Note that $\sigma^2_{\mathrm{photometric}}$ is defined as a constant for all sample points. It has two components: the unaccounted-for uncertainties, such as the reddening or crowding effects and the non-negligible width of the instability strip (the intrinsic scatter). For the fitting, we assumed flat priors for all parameters. The resulting fit along with the derived parameters is shown in Fig.~\ref{fig:N4258}. 

To estimate the distance of M~51, we applied the result obtained for NGC~4258 by fitting Eq.~\ref{eq:PL}: we fixed the slope of the PL relation to the one obtained from the reference fit and measured its offset relative to that of the anchor galaxy. We determined the metallicity of each M~51 Cepheid based on its position within the host and by adopting the metallicity gradient obtained by \cite{Zaritsky94}. According to this, on average, M~51 has 0.4 dex higher metallicity than the anchor galaxy. To determine the distance of M~51 using only NGC~4258 as anchor, we measured the M~51 metallicities relative to those of the NGC~4258 Cepheids, hence it was not necessary to assume a reference solar value.

Several overtone Cepheids are present in the M~51 set, which can be used for the distance determination as well. However, making use of them requires the application of a few changes compared to the regular PL analysis, hence we conducted two versions of the Cepheid-based distance estimation to M51: one without and one with overtone Cepheids.

\subsubsection{Fundamental Mode only fit}
In the 'fundamental mode only' version, we have applied a period cut of 10 days to separate the fundamental mode Cepheids from their overtone counterparts. The period limit for this cut was motivated by the photometric incompleteness, as fundamental mode Cepheids with shorter periods were too faint and were not present in our sample. Hence, the majority of the removed stars were overtone Cepheids. To fit the PL relation to the filtered dataset, we applied the outlier rejection method presented in \cite{Kodric2015, Kodric2018}. This method works iteratively, and it is based on the median absolute deviation (MAD) of the data points. 

Throughout this iterative procedure, at each step, the MAD of the dataset is calculated, and then the data point with the greatest deviation is discarded, as long as it is at least $\kappa$ times away from the model value (where $\kappa$ is a tuning parameter). After this, the MAD is recalculated, and the rejection criterion is re-evaluated. For our work, we adopted $\kappa = 4$ in line with the discussion in \cite{Kodric2015}. The advantage of the method is that it removes the outliers one by one, which is not only controllable but it is also governed by the statistics of the residuals, hence no arbitrary cuts have to be made. 

\begin{figure}
    \centering
    \includegraphics[width = \linewidth]{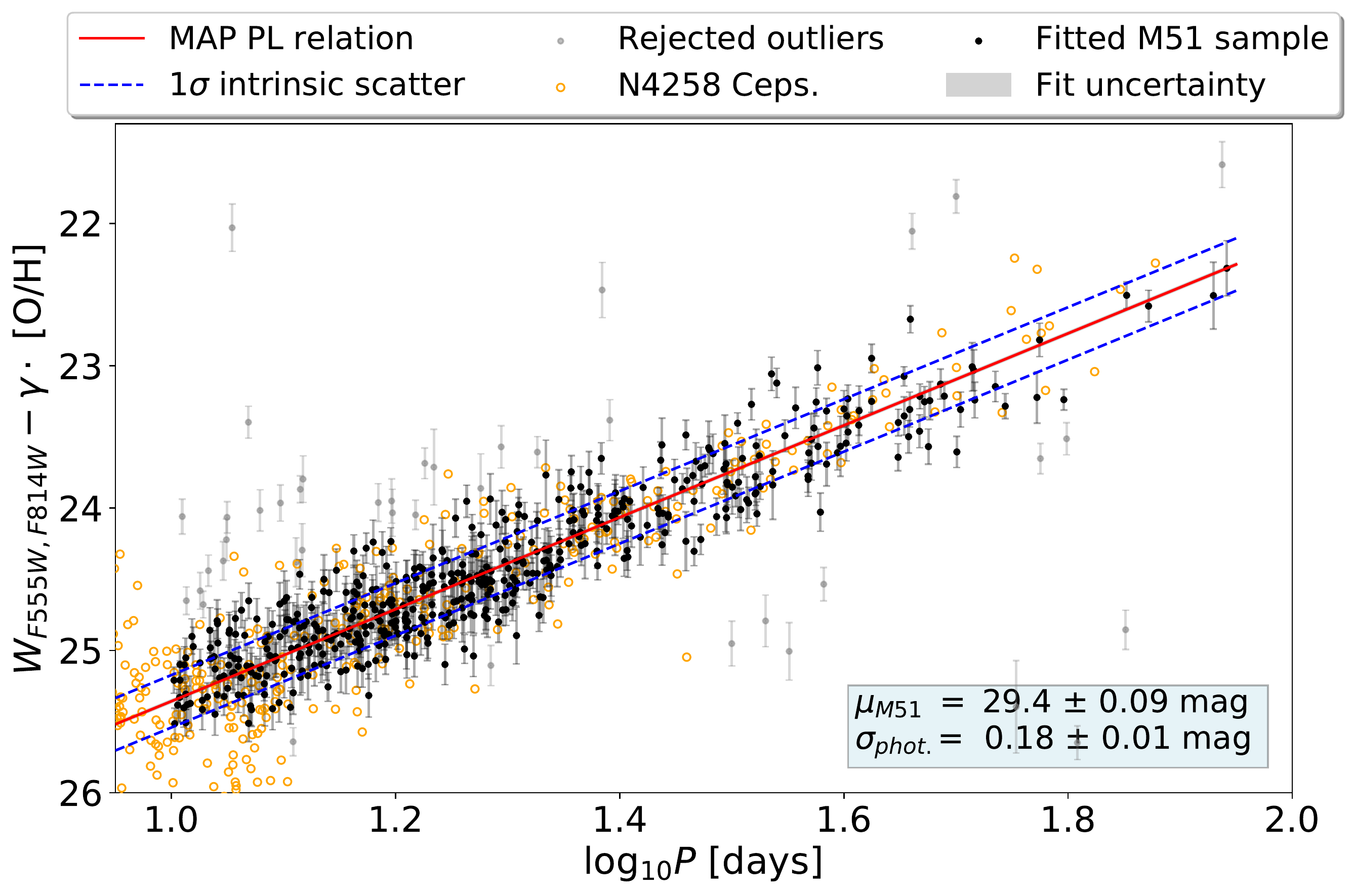}
    \caption{The PL relation fitting of the fundamental mode Cepheids in M~51, above the period limit of $P > 10$ days. The yellow open circles show the NGC~4258 Cepheids for reference, while the faint grey points denote the stars which were rejected by the outlier detection method. The red line corresponds to the Maximum A Posteriori (MAP) estimate, i.e. the most likely fit. The displayed distance uncertainty includes the systematic terms (namely the uncertainty in the metallicity correction and the reference NGC~4258 PL relation).}
    \label{fig:M51-PL-1}
\end{figure}

The obtained fit for the PL relation of the M~51 Cepheids in this setup is displayed in Fig.~\ref{fig:M51-PL-1}. To fit the M~51 sample, we have recalculated the fit parameters for the anchor galaxy NGC~4258 using only the $P > 10 $ day period range, to obtain an unbiased estimate on the slope (although the resulting anchor slope and offset after the step-by-step outlier removal and fitting were matching the original values to better than $1\%$). This way, we have calculated a distance modulus of $\mu_{M51} = 29.40 + 0.09$ mag, and a distance estimate of $D = 7.59 \pm 0.30$ Mpc for M~51.

It is worth comparing the fitted observed scatter values: for M~51 we obtained $\sigma_{\textrm{phot., M51}} = 0.20$ mag, while for NGC~4258 $\sigma_{\textrm{phot., N4258}} = 0.16$ mag. A similar result of $\sigma_{\textrm{phot., M101}} = 0.21$ mag can be obtained for M~101 based on the \cite{Hoffmann2016} data. It is important to note, that for the comparison NGC~4258 and M~101 values F160W data were also used, which generally reduce the scatter of the Wesenheit magnitudes. The similarity of these values shows well the precision of the C18 data and the high quality of the Cepheid sample.

\subsubsection{Fundamental Mode + First Overtone fit}
To investigate how the presence of overtone Cepheids changes the distance estimate, we attempted to fit the sample without removing the overtone variables. For this estimation, we assumed that the PL relation slopes of the fundamental mode and overtone Cepheids are the same, and match that of the NGC~4258 Cepheid PL relation. However, it is not known exactly how much brighter the overtone Cepheids should be in the chosen Wesenheit system. Although in principle distinguishing the overtone Cepheids from the fundamental mode ones can be done based on light curves either due to their low number they were not separated by PCA. The analysis of the Fourier components also did not separate these two subtypes, due to the relatively high photometric errors on these Cepheids. We thus attempted to separate these two types statistically, based on their period-magnitude values.

To do this, we introduced two criteria during the fit. First, we assumed that all stars with $P > 10$ days are fundamental mode Cepheids, since overtone Cepheids of this period are unlikely \citep{Baranowski2009}. For the second criterion, we introduced an offset parameter $\Delta$, which measured the magnitude difference between the fundamental mode and overtone PL relations. Throughout the fitting, each star below the period limit that was at least $1/2 \cdot \Delta$ magnitudes brighter than the fundamental mode PL relation was assigned to the overtone class. Otherwise, it stayed among the fundamental mode variables. This way, we could fit the PL relation of both modes simultaneously. This ad-hoc classification was revised every time a new $\Delta$ value was chosen. However, since the Bayesian fitting of this model turned out to be infeasible (due to the simultaneous incorporation of the intrinsic scatter and the offset of overtone Cepheids), we chose not to fit but to marginalize over this offset.

\begin{figure}
    \centering
    \includegraphics[width = \linewidth]{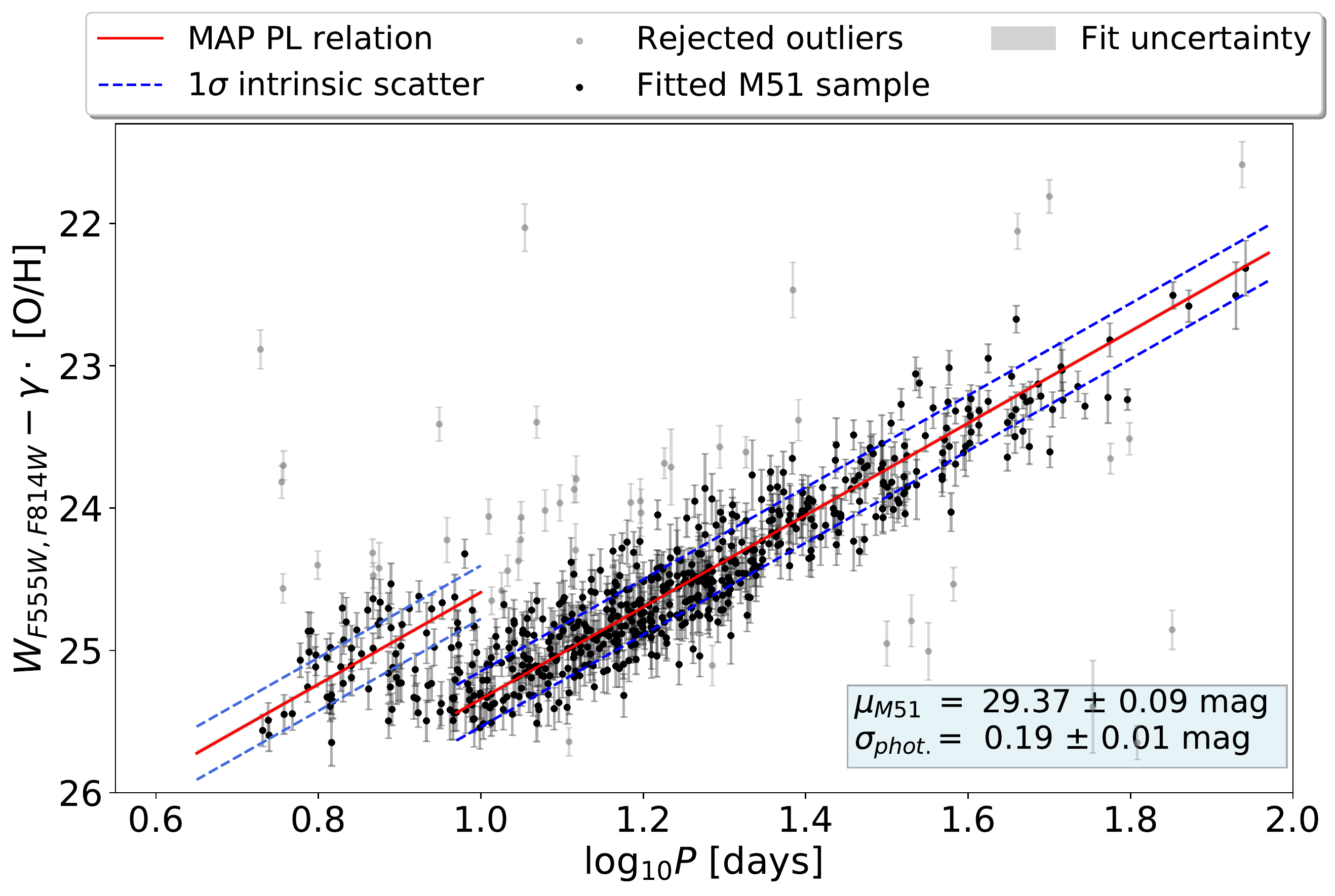}
    \caption{The simultaneous fitting of both fundamental mode and overtone PL relation for the M~51 Cepheid sample. For this run, an offset of 0.75 magnitudes was assumed between the two modes. The red line corresponds to the Maximum A Posteriori (MAP) estimate. The light grey points denote the data points that were rejected by the outlier detection method, while the solid points show the ones that were used for the fitting of the relation. As before, the displayed distance uncertainty includes the aforementioned systematic terms.}
    \label{fig:M51-over}
\end{figure}

To marginalize, we have set up multiple fits assuming different values from a reasonable offset range of [0.45, 0.95] mag. We ran the fitting for all these multiple setups (one of which is shown in Fig.~\ref{fig:M51-over}), including the previously described outlier rejection, yielding a distance posterior each. Then, we combined these distributions to obtain a marginalized posterior, which included the uncertainty in the offset of the fundamental mode -- overtone PL relations.

\begin{figure}
    \centering
    \includegraphics[width = \linewidth]{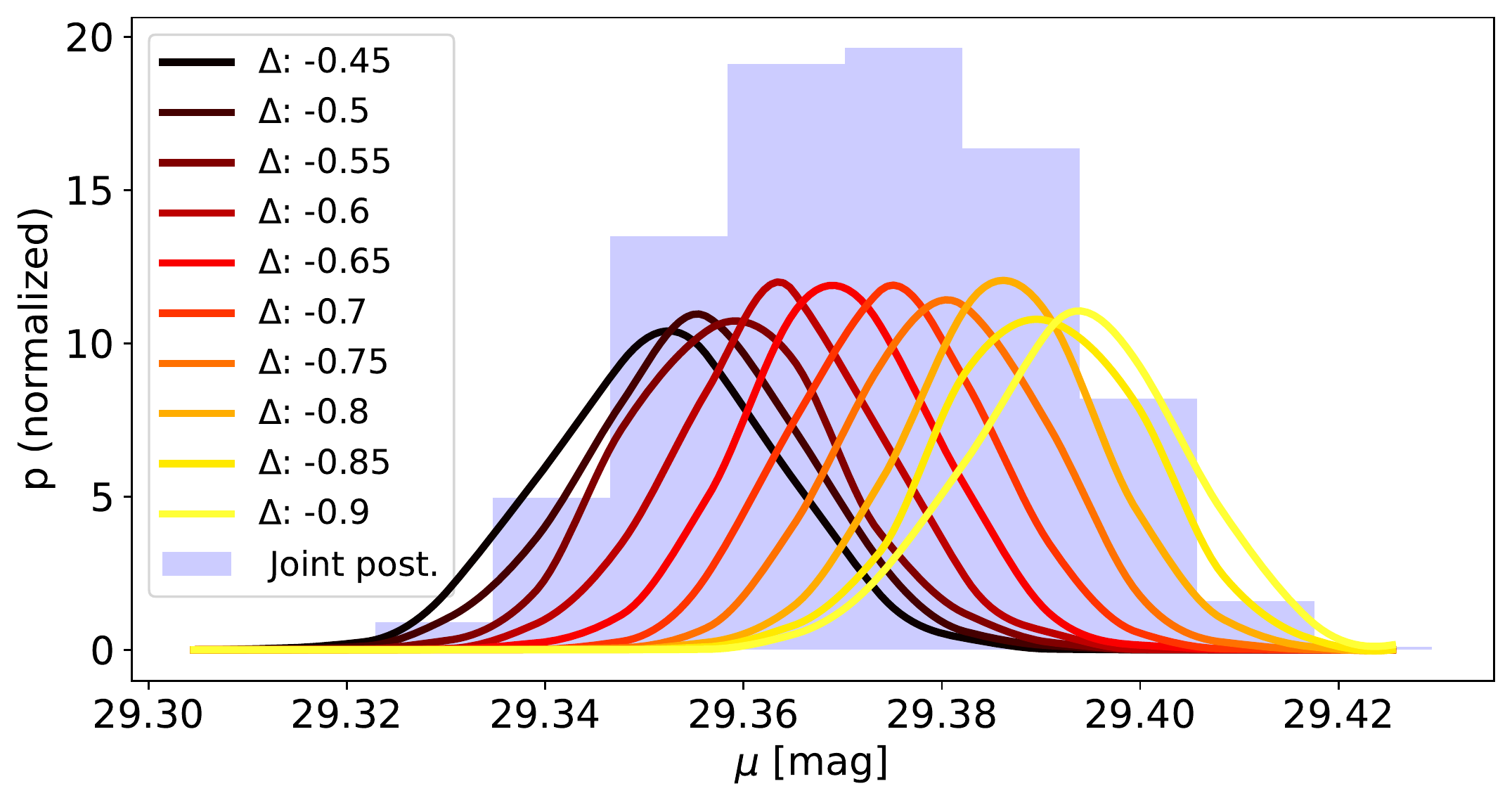}
    \caption{The individual distance posteriors obtained for different offset values $\Delta$ (solid curves) and the combined posterior (histogram in the background).}
    \label{fig:all_post}
\end{figure}

Fig.~\ref{fig:all_post} shows the combined posterior. By averaging over the combined posterior and then propagating the PL relation fitting and distance modulus uncertainties valid for NGC~4258, we calculated a value of $\mu_{M51} = 29.37 \pm 0.09$ mag (a relative distance modulus of $\mu_{M51} - \mu_{N4258} = 0.03 \pm 0.09$ mag), which corresponds to a distance estimate of $D = 7.49 \pm 0.30$ Mpc for M~51. This estimate is consistent with the distance calculated based on the fundamental mode Cepheids only. 

To cross-check the precision of our NGC 4258-based calibration, we also carried out the distance measurement utilizing Milky Way open cluster Cepheids as an anchor. For this, we used the recent work of \cite{CruzReyesAnderson2023}, in which the period-luminosity relation of Milky Way Cepheids was refined. To perform the calibration, we estimated the $\alpha$ and $\beta$ fit parameters using equations 26. and 27. presented in \cite{CruzReyesAnderson2023} and based on the pivot wavelengths of the relevant F555W and F814W bands (adopted from \url{http://svo2.cab.inta-csic.es/svo/theory/fps3/}). The resulting calibration parameters read $\alpha = -3.471$ and $\beta = -5.998$ (the latter of which contains the metallicity term, $\gamma \cdot $[O/H]$_{\textrm{MW,Cep}} \sim -1.76$). To account for the metallicity difference, we measured the [O/H] values relative to that of the Milky Way Cepheid sample when fitting Eq.~\ref{eq:PL}. By carrying out the same distance measurement procedure as above, we arrived at distances of $\mu = 29.45 \pm 0.12$ mag and $\mu = 29.43 \pm 0.12$ mag for the 'fundamental mode only' and the 'fundamental mode and first overtone' versions respectively, which are in perfect agreement with the results obtained previously. This validates well our NGC~4258-based calibration.

\section{SN~2005cs}
\label{sec:2005cs}
SN~2005cs is one of the best-known SNe IIP in the literature, mostly owing to the fact that it belongs to the subclass of peculiar underluminous objects \citep{Pastorello2009, Kozyreva2022}. Given that the host is a well-observed target by amateur astronomers, several non-detections were available before its explosion, the latest one just a day before the first detection. The photometric and spectroscopic observations of this supernova are extensively described in \cite{Pastorello2009}. The data we adopted and used for this supernova are summarized in Sect.~\ref{sec:data}. The time series of this supernova was analysed several times with the purpose of obtaining a distance to it based on the standardisable candle method \citep{Hamuy2005} and the expanding photosphere method \citep{Takats2006, Dessart2008, Vinko2012, Takats2012}. The latest independent EPM analysis by \cite{Vinko2012} yielded a distance of $8.4 \pm 0.7$ Mpc based on photospheric velocity measurements using model spectra generated by SYNOW \citep{synow}.

We re-performed the tailored-EPM-based analysis of SN~2005cs using the spectral emulator introduced in \cite{Vogl2020}. The main goal of reanalysing the data of SN~2005cs is twofold: on the one hand, we aim to investigate how the improvements of the spectral fitting method influence the outcome of the analysis, while on the other hand, we wish to compare this updated result to the independently obtained Cepheid distance. We stress that we do not aim to calibrate the SN II method based on this Cepheid distance since the EPM does not require any such calibration; instead, we are performing a single object consistency test for the two methods. The EPM we are using was already showcased on SN~2005cs in \cite{Vogl2020}; here, we discuss the extended version of this analysis (which includes a more complete constraint for the time of explosion, the calibration of spectra to contemporaneous photometry and the more advanced treatment of reddening in the EPM regression). These improvements have been discussed in detail in \cite{Csornyei2022}.

\subsection{Time of explosion}
Estimating a high-quality distance to a supernova based on the expanding photosphere method requires precise knowledge of the time of explosion. Often, this parameter is determined as the mid-point between the first detection and the last non-detection, with an assumed uncertainty of half of the time elapsed between the two. However, this does not make use of all available information, such as the rise of the light curve. Taking this into account one can improve the EPM results by constraining the time of explosion with higher precision. Henceforth, we determined the time of explosion based on fitting the early light curve by an inverse exponential following the reasoning of \cite{Ofek2014} and \cite{Rubin2016}. We fitted the flux $f$ in band $W$ with a model 

\begin{equation}
    f_W(t) = f_{m, W} \left[1 - \exp{\left(-\frac{t-t_0}{t_{e, W}}\right)}\right],
\end{equation}

\noindent where $t$ is the time, $t_0$ is the time of explosion, $f_{m, w}$ is the peak flux, and $t_{e,W}$ is the characteristic rise time in the particular band. We carried out this fitting for multiple photometric bands simultaneously to increase the accuracy of the method. Similarly, as in \cite{Csornyei2022}, $t_0$ was treated as a global parameter for the fitting (i.e. it was the same for all bands), while each of the different bands had their own $f_{m,W}$ and $t_{e,W}$ parameters. We imposed the additional constraint on the joint fit that the characteristic rise time should increase with wavelength as seen in well-observed SNe before \citep[see, e.g.,][]{Gonzalez2015}. 

\begin{figure}
    \centering
    \includegraphics[width=\linewidth]{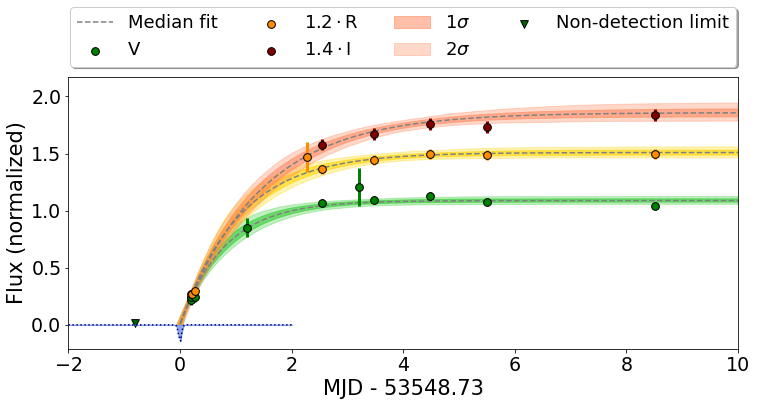}
    \caption{Exponential fit to the light curves of SN 2005cs, with the amateur observations included. The shaded regions denote the 68$\%$ and 95$\%$ confidence intervals.}
    \label{fig:2005cs_elc_mod}
\end{figure}

Given that SN~2005cs was observed very early on owing to the regular amateur observations of M~51 (see \citealt{Pastorello2009} for a list of these observations, one of which is well described at \url{https://birtwhistle.org.uk/GallerySN2005cs.htm}), the time of explosion can be very tightly constrained through the exponential fitting. Performing the fitting including the amateur data yielded a precise $t_0$ estimate of JD $2453549.23^{+0.03}_{-0.03}$ (see Fig.~\ref{fig:2005cs_elc_mod}). This estimate was then used as an independent prior for the EPM regression.



\begin{figure*}
    \centering
    \includegraphics[width=0.495\linewidth]{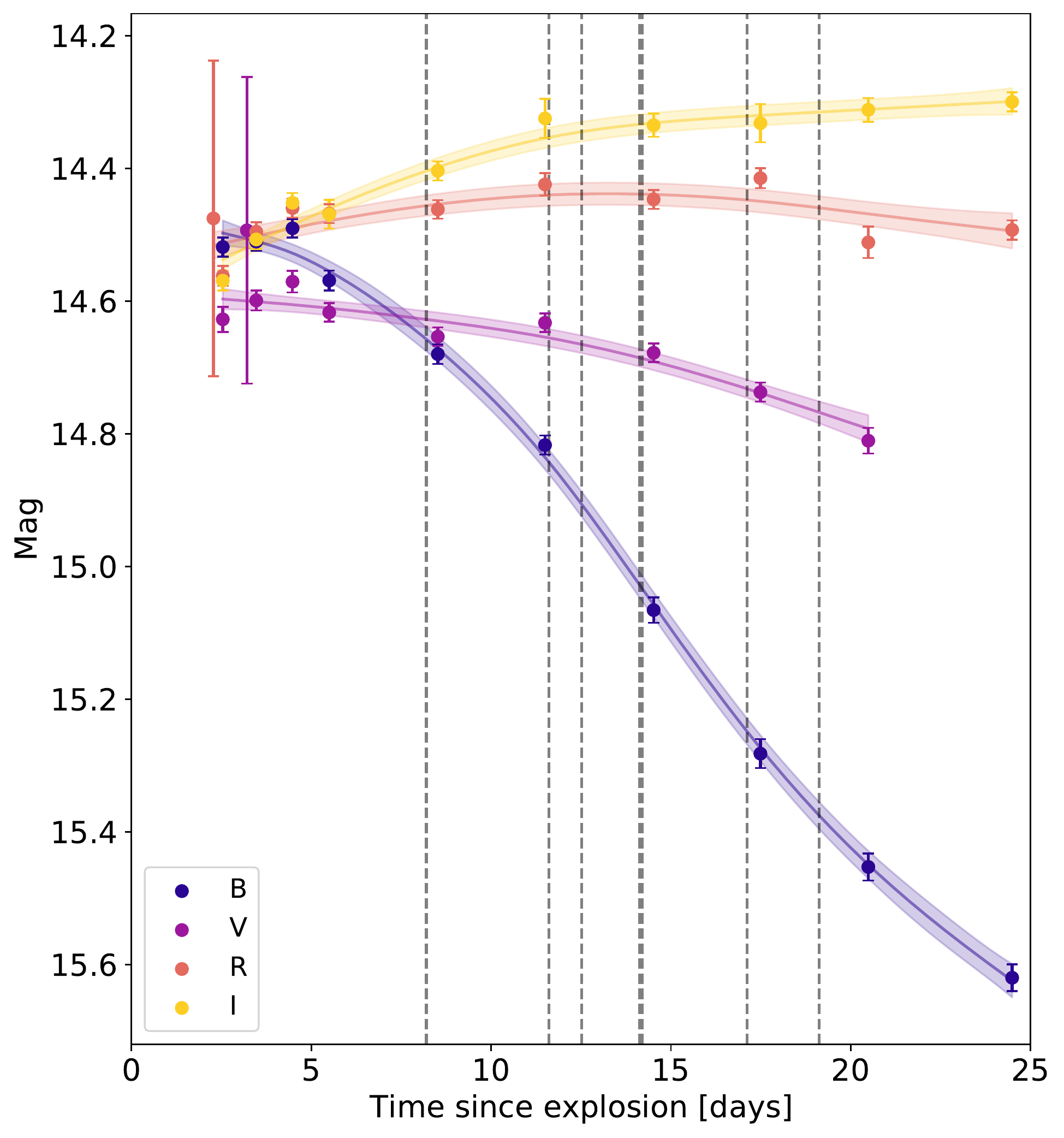}
    \includegraphics[width=0.485\linewidth]{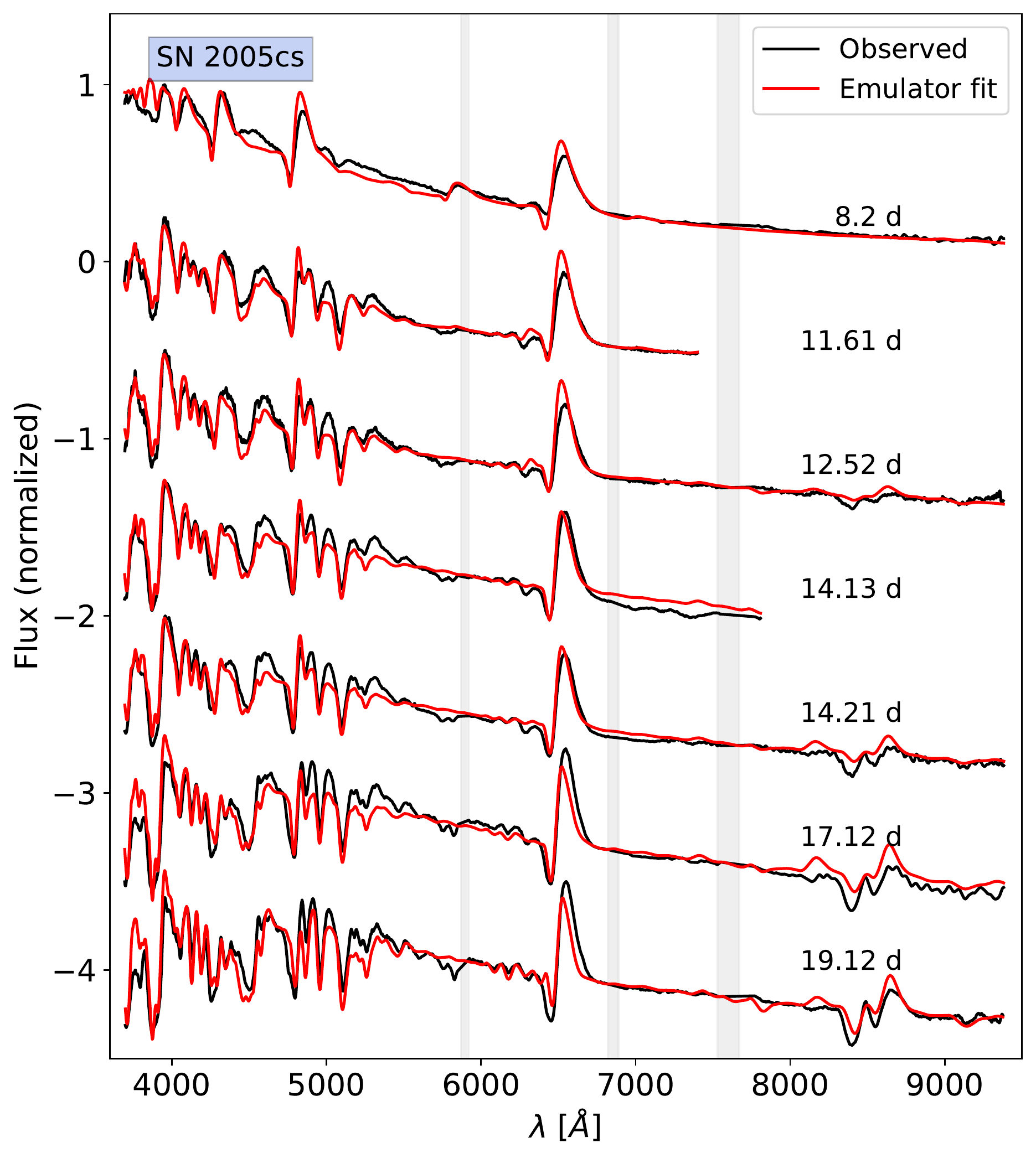}
    \caption{\textbf{Left:} Gaussian-Process light curve fits for 2005cs in the various bands. The grey dashed lines denote the epochs at which a spectrum was taken that was included in our sample. The epochs are measured with respect to MJD 53548.73. \textbf{Right:} The spectra and their emulator fits for the various epochs for an assumed reddening of $E(B-V) = 0.03$ mag. The grey bands indicate the telluric regions and the sodium band that were masked for the fitting.}
    \label{fig:fits}
\end{figure*}

\subsection{Interpolated light curves}
To measure the distance to SN~2005cs we require knowledge of the absolute and observed magnitudes at matching epochs, which are then compared through Eq.~\ref{eq:EPM_mod} and Eq.~\ref{eq:EPM}. Determining the former requires the modelling of spectral observations, however, photometry is rarely available simultaneously with spectral epochs. We calculated the observed magnitudes at these epochs by reconstructing the light curves using Gaussian Processes (GPs). For the implementation, we utilized the \texttt{george}\footnote{\url{https://george.readthedocs.io/en/latest/}} python package \citep{george}. GPs present an excellent way to interpolate light curves, as they provide a non-parametric way of fitting, while taking into account the uncertainties in the data. As a result, we could obtain smooth and continuous light curve fits (see Fig.~\ref{fig:fits}), which could be used for brightness estimation at the spectral epochs. The interpolated magnitudes are listed in Table~\ref{tab:photometry}.

\subsection{Flux calibration of spectra}
Determining the absolute magnitudes for EPM requires estimations of the physical parameters of the supernova, which are best determined through spectral fitting. Since the emulator-based modelling and hence the tailored-EPM analysis requires well-calibrated spectral time series (which is also a requirement for the precise determination of the extinction), the individual spectra had to be re-calibrated first based on the photometry. We calculated the relevant set of synthetic magnitudes using the response curves from \cite{Bessell2012} and compared these to the corresponding interpolated magnitudes. To correct for flux calibration differences that can be approximated as linear in wavelength, we fitted the first-order trend present in the pairwise ratios of synthetic and interpolated magnitudes against the effective wavelengths of the passbands (including uncertainty inflation following the description of \citealt{Hogg2010}). We then corrected the spectra for this trend.


\begin{table}[]
    \centering
    \begin{tabular}{c c c c c}
    \toprule
    Epoch [d] & $B$ & $V$ & $R$ & $I$\\
    \hline
8.20  & 14.65 & 14.62 & 14.45 & 14.40\\
11.61 & 14.84 & 14.65 & 14.43 & 14.35\\
12.52 & 14.94 & 14.67 & 14.43 & 14.33\\
14.13 & 15.02 & 14.68 & 14.43 & 14.33\\
14.21 & 15.03 & 14.68 & 14.43 & 14.33\\
17.12 & 15.24 & 14.73 & 14.44 & 14.32\\
19.12 & 15.37 & 14.76 & 14.46 & 14.31\\
    \hline
    \end{tabular}
    \caption{Interpolated magnitudes for SN~2005cs. The epochs are measured with respect to the estimated time of explosion MJD 53548.73}.
    \label{tab:photometry}
\end{table}

\subsection{Spectral modelling and tailored-EPM}
\label{sec:specmod}
To fit the individual re-calibrated spectra, we applied the methodology from \cite{Vogl2020} and passed the spectral time series to the emulator. This emulator allows for a fast and reliable interpolation of simulatetd spectra for a given set of physical parameters. The parameter space of the training sets used for the emulator is summarized in Table~\ref{tab:emulator}. These training sets are identical to the ones used in \cite{Csornyei2022}. Similarly to that work, the choice of the training set used for the emulator is based on the epoch of the spectrum: normally, one would use the first set for spectra younger than 16 days, and the second set for older ones. However, due to the accelerated evolution of SN~2005cs, the training set designed for the modelling of primarily older spectra had to be employed already at an epoch of 10 days.

\begin{table}[t]
  \centering
  \resizebox{\columnwidth}{!}{
  \begin{tabular}{cccccccc}
   \toprule \toprule
     & $v_\mathrm{ph}$\,[km$\,\mathrm{s}^{-1}$]  & $T_\mathrm{ph}\,$[K] & $Z\,[Z_\odot]$ & $t_\mathrm{exp}\,$[days] & n & \multicolumn{2}{c}{NLTE} \\
    \midrule
    \multicolumn{6}{c}{$\mathbf{t_\mathrm{exp}\,}$ \textbf{< 10 days}} & \textbf{\small{H}} & \textbf{\small{He}} \\
    \midrule
    Min & 4500 & 7200  & 0.1 & 2.0 & 9 & \multirow{2}{*}{\cmark} & \multirow{2}{*}{\cmark}  \\
    Max & 12000 &  16000 & 3.0 & 16.0 & 26 \\
    \midrule
    \midrule
    \multicolumn{6}{c}{$\mathbf{t_\mathrm{exp}\,}$ \textbf{> 10 days}} & \textbf{\small{H}} & \textbf{\small{He}} \\
    \midrule
    Min & 3600 &  5800 & 0.1 & 6.5 & 6 & \multirow{2}{*}{\cmark} & \multirow{2}{*}{\xmark} \\
    Max & 10700 &  10000 & 3.0 & 40.0 & 16 \\
    \bottomrule
  \end{tabular}
  }
  \caption[Parameter range covered by the extended spectral emulator]{Parameter range covered by the extended spectral emulator. The individual columns show the various physical parameters: $v_{\textrm{ph}}$ and $T_{\textrm{ph}}$ denote the photospheric velocity and temperature, $Z$ the metallicity, $t_{\textrm{exp}}$ the time since explosion and $n$ the exponent of the power-law density profile. The last columns show whether an NLTE (non-local thermal equilibrium) treatment was also included for H and He. The two table sections correspond to the two training sets as described in the main text.}
  \label{tab:emulator}
\end{table}

The epoch $t$ of each spectrum was fixed. While the physical parameters were estimated by the emulator directly, we treated the reddening separately: as in \cite{Csornyei2022}, we set up a grid of possible $E(B-V)$ values and performed the maximum likelihood fitting for each of them by reddening the synthetic spectra with the given value. For the lower limit of the $E(B-V)$ grid, we assumed the Galactic colour excess towards the supernova, which was determined based on the \cite{Schlafly2011} dust map. The best fit $E(B-V)$ was then chosen as the average of the $E(B-V)$ values that resulted in the lowest $\chi^2$ for the individual spectra.

\begin{table*}[]
    \centering
    \begin{tabular}{c c c c c c c c c c}
    Epoch [d] & $T_{\textrm{ph}}$ [K] & $v_{\textrm{ph}}$ [$\frac{\textrm{km}}{\textrm{s}}$] & $n$ & $\Theta$ [$10^8$ $\frac{\textrm{km}}{\textrm{Mpc}}$] & $\frac{\Theta}{v}$ [$\frac{\textrm{d}}{\textrm{Mpc}}$] & $T_{\textrm{C20}}$ [K]& $v_{\textrm{C20}}$ [$\frac{\textrm{km}}{\textrm{s}}$] & n$_{\textrm{C20}}$ & $\frac{\Theta}{v}_{\textrm{C20}}$ [$\frac{\textrm{d}}{\textrm{Mpc}}$] \\
    \hline
    8.20 & 9364 & 5620 & 9.80 & 5.107 & 1.052 & - & - & - & - \\
    11.61 & 7106 & 4816 & 11.16 & 6.617 & 1.590 & 7003 & 4766 & 12.2 & 1.69 \\
    12.52 & 6815 & 4509 & 11.60 & 6.821 & 1.736 & 6799 & 4502 & 11.8 & 1.90\\
    14.13 & 6176 & 4354 & 13.72 & 7.553 & 2.008 & - & - & - & -\\
    14.21 & 6279 & 4223 & 13.47 & 7.435 & 2.038 & 6720 & 4086 & 10.9 & 2.12\\
    17.12 & 6001 & 3993 & 14.13 & 7.650 & 2.218 & 6457 & 4499 & 12.3 & 2.01\\
    19.12 & 5998 & 3694 & 13.63 & 7.819 & 2.450 & 6349 & 3721 & 12.4 & 2.44\\
    \hline
    \end{tabular}
    \caption{Table of the inferred physical parameters for SN~2005cs. The $\Theta$ values were calculated for the $BVI$ bandpass combination. The values with the $\textrm{C20}$ subscript refer to the values found by \cite{Vogl2020} using an earlier version of the emulator. The epochs are measured with respect to the estimated time of explosion MJD 53548.73.}
    \label{tab:fit_pms}
\end{table*}

Since SN~2005cs was a low luminosity supernova, its spectral evolution was accelerated compared to other normal Type IIP supernovae \citep{Pastorello2009}. Due to this and its lower temperatures, it reached the limits of the emulator faster. Hence we could only use spectra from the first 20 days for the tailored-EPM. The re-calibrated spectral sequence of SN~2005cs and the corresponding models for the best-fit reddening are shown in Fig.~\ref{fig:fits}. The best-fit parameters for the individual epochs are summarized in Table~\ref{tab:fit_pms} along with with the previous results from \cite{Vogl2020}. The fits favoured minimal reddening, hence we adopted the reddening of $E(B-V) = 0.03$ mag obtained by \cite{Schlafly2011} towards M~51. This is similar to the estimate that can be obtained from the NaID features seen in the spectra of its sibling supernova, SN~2011dh \citep{Vinko2012}, hence it is unlikely that the foreground extinction towards M~51 is higher. Moreover, our best-fit extinction estimate is consistent with the value found by previous studies on the spectra of SN~2005cs \citep{Baron2007, Dessart2008}.

\begin{figure*}
   \centering
    \includegraphics[width=\linewidth]{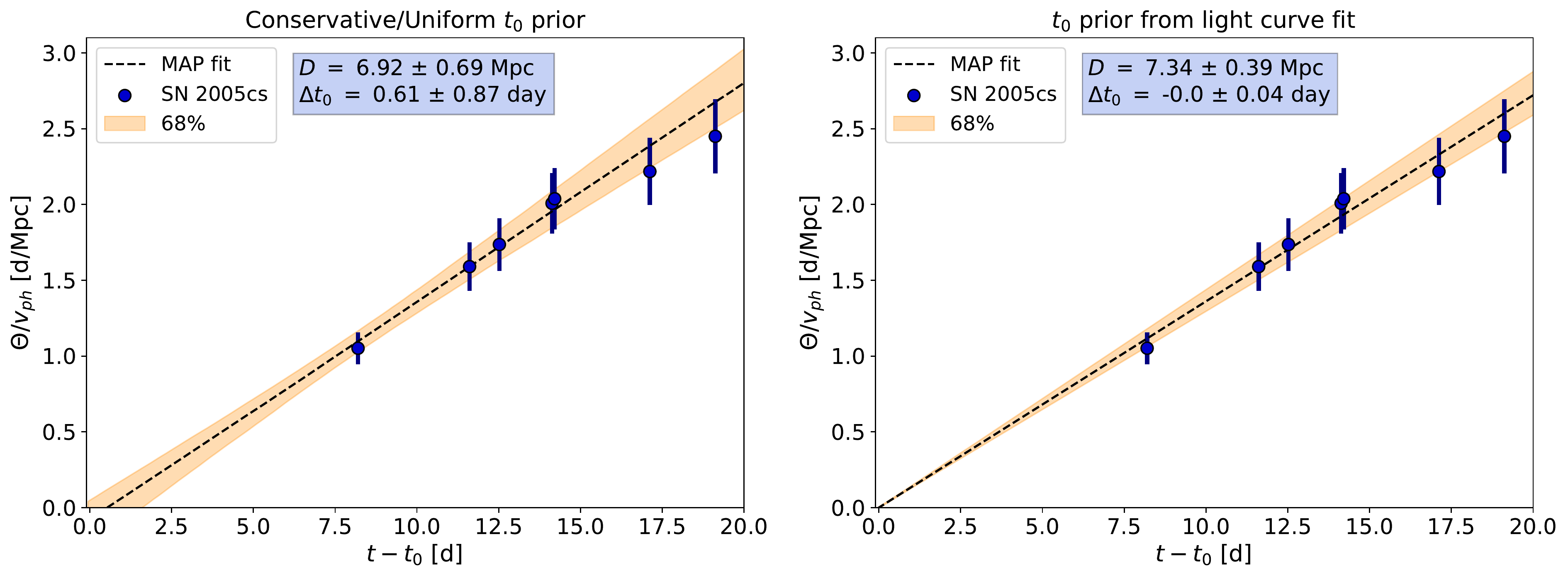}
    \caption{The EPM regression obtained for SN~2005cs in the two versions (the \textbf{left} plot shows the more conservative approach, where we adopted a flat prior for $t_0$, while the \textbf{right} shows the fit for the $t_0$ prior that is informed by the light curve). The $x$ axis shows the time elapsed since the explosion epoch $t_0$, MJD 53548.73 for both plots. The points show the evolution of $\Theta / v_{\textrm{ph}}$ for this supernova, inferred from spectral fitting. The shaded region shows the uncertainty of the fit.}
    \label{fig:2005cs_EPM}
\end{figure*}

 In general, we find that the obtained $\Theta/v$ estimates are in good agreement with the previous values of \cite{Dessart2008}. We note that \cite{Dessart2008} modelled two earlier epochs of SN~2005cs as well (at epochs of 4 and 5 days), however, they pointed out that the emission line profiles could not be adequately reproduced. When modelling these epochs, we encountered similar issues and found the estimated physical parameters to be unreliable. Due to these shortcomings, we decided not to include these epochs in the analysis.
 
 As a final step of the distance determination, we performed the EPM analysis of SN~2005cs, by fitting a linear function to the $\Theta / v_{ph} (t)$ values using \texttt{UltraNest}. We performed two fits with different priors on the time of explosion. For the first version, we picked a uniform prior in the range of [$-1.3$,$2.5$] days around the assumed $t_0$ value which is set by the last non-detection (0.5 day before the non-detection, to remain conservative) and the first KAIT detection. For the second one, we set the prior based on the fit shown in Fig.~\ref{fig:2005cs_elc_mod}. For both fits, the prior on the distance was set to be flat. To include the systematic uncertainties caused by the reddening in the final error estimate, we applied the treatment described in \cite{Csornyei2022} for both versions.

The results of both EPM regressions are shown in Fig.~\ref{fig:2005cs_EPM}. In the more conservative case, when the amateur observations are not taken into account, we obtain a distance of $D = 6.92 \pm 0.69$ Mpc, which corresponds to $\mu = 29.20 \pm 0.20$ mag. In case we tighten the $t_0$ prior used for the EPM regression based on the light curve, the final fit yields a distance of $D = 7.34 \pm 0.39$ Mpc (i.e. $\mu = 29.33 \pm 0.11$ mag). We point out that the resulting $t_0$ estimate of the conservative version is in tension with the amateur photometric data, as SN~2005cs was already visible on images taken earlier. Hence, we strongly favour the approach which makes use of the early light curve fit. We also note that including the tighter constraints reduces the EPM distance uncertainties almost by a factor of two. This highlights the importance of having an informed prior on the time of explosion for a precise distance estimate.

\section{Discussion}
\label{sec:discussion}

We measured two independent distances for M~51: $D = 7.59 \pm 0.30$ Mpc based on Cepheid variables and the period-luminosity relation (by including fundamental mode Cepheids only, otherwise this modifies to $D = 7.49 \pm 0.30$ Mpc when using both fundamental mode and first overtone Cepheids and applying the alternative fitting), and $D = 7.34 \pm 0.39$ Mpc based on the updated EPM modelling of SN~2005cs ($D = 6.92 \pm 0.69$ Mpc in the case of the more conservative approach). This consistency highlights the precision of the tailored-EPM along with its robustness, while it strengthens the Cepheid distance as well. Given that the Cepheid and SN 2005cs-based estimates are completely independent, one can take their average to obtain a higher-precision distance; combining the fundamental mode Cepheid only distance with the favoured SN~2005cs estimate, the resulting value is $D = 7.50 \pm 0.24$ Mpc. This estimate is precise to 3.2\%, which is remarkable for an extragalactic distance. As we will expand on later, this distance is more than $10\%$ lower than the estimates used previously for this galaxy.


It is important to note for the Cepheid distances, that in our analysis we did not consider the effects of the stellar association bias (i.e. Cepheids may be blended with their birth clusters, which would cause a bias in the estimated brightness). The effect of this bias on the estimated distances was investigated by \cite{AndersonRiess2018} and will be further inspected by \cite{Spetsieri_prep}. Based on their results, accounting for the bias can have sub-percent effects on the Cepheid brightnesses, and make the inferred distance larger by the same amount (on the scale of $10^{-3}$ mag, as noted in \citealt{AndersonRiess2018}). We note that by accounting for this bias, our Cepheid distance would match the supernova-based result slightly worse, but they would still remain fully consistent.

\begin{table}[]
    \centering
    \scalebox{0.8}{
    \begin{tabular}{c c c c }
    \hline
    $\mu$ [mag] & $D$ [Mpc] & Method & Data \\
    \hline
    \multicolumn{4}{c}{\textbf{This work}}\\
    $\mathbf{29.40 \pm 0.09}$ & $\mathbf{7.59 \pm 0.30}$ & Fund. mode PL & \cite{Conroy2018}\\
    $\mathbf{29.37 \pm 0.09}$ & $\mathbf{7.49 \pm 0.30}$ & Fund. + Overtone PL & \cite{Conroy2018}\\
    $\mathbf{29.20 \pm 0.20}$ & $\mathbf{6.92 \pm 0.69}$ & SN~2005cs EPM flat $t_0$ & 2005cs\\
    $\mathbf{29.33 \pm 0.11}$ & $\mathbf{7.34 \pm 0.39}$ & SN~2005cs EPM LC $t_0$ & 2005cs\\
    \hline
    \hline
    $\mu$ [mag] & $D$ [Mpc] & Reference & Data \\
    \hline
    \multicolumn{4}{c}{\textbf{Tip of the Red Giant Branch (TRGB)}}\\
    $29.67 \pm 0.09$ & $8.58 \pm 0.36$ & \cite{McQuinn2017} & \cite{McQuinn2016} \\ 
    $29.78 \pm 0.13$ & $9.05 \pm 0.54$ & \cite{Tikhonov2015} & archival HST\\
    $29.79 \pm 0.14$ & $9.09 \pm 0.59$ & \cite{Tikhonov2015} & archival HST\\
    $29.74 \pm 0.14$ & $8.88 \pm 0.57$ & \cite{Tikhonov2015} & archival HST\\
    \hline
    \multicolumn{4}{c}{\textbf{SN optical}}\\
    $29.46 \pm 0.12$ & $7.80 \pm 0.43$ & \cite{Vogl2020} & 2005cs \\ 
    $29.46 \pm 0.11$ & $7.80 \pm 0.40$ & \cite{Pejcha2015} & 2005cs \\
    $29.77 \pm 0.08$ & $8.99 \pm 0.33$ & \cite{Rodriguez2014} & 2005cs \\
    $29.63 \pm 0.05$ & $8.43 \pm 0.19$ & \cite{Rodriguez2014} & 2005cs \\
    $29.51 \pm 0.14$ & $7.97 \pm 0.51$ & \cite{Bose2014} & 2005cs\\
    $29.37 \pm 0.04$ & $7.49 \pm 0.14$ & \cite{Bose2014} & 2005cs\\
    $28.96 \pm 0.17$ & $6.20 \pm 0.48$ & \cite{Bose2014} & 2005cs\\
    $28.91 \pm 0.05$ & $6.06 \pm 0.14$ & \cite{Bose2014} & 2005cs\\
    $29.62 \pm 0.05$ & $8.40 \pm 0.19$ & \cite{Vinko2012} & 2005cs, 2011dh\\
    $29.67 \pm 0.05$ & $8.60 \pm 0.20$ & \cite{Takats2012} & 2005cs\\
    $29.38 \pm 0.06$ & $7.50 \pm 0.21$ & \cite{Takats2012} & 2005cs\\
    $29.61 \pm 0.21$ & $8.35 \pm 0.81$ & \cite{Poznanski2009} & 2005cs\\
    $29.75 \pm 0.12$ & $8.90 \pm 0.49$ & \cite{Dessart2008} & 2005cs\\
    $29.75 \pm 0.16$ & $8.90 \pm 0.66$ & \cite{Dessart2008} & 2005cs\\
    $29.50 \pm 0.18$ & $7.90 \pm 0.66$ & \cite{Baron2007} & 2005cs\\
    $29.40 \pm 0.29$ & $7.59 \pm 1.01$ & \cite{Takats2006}& 2005cs\\
    $29.02 \pm 0.44$ & $6.36 \pm 1.29$ & \cite{Takats2006}& 2005cs \\
    $29.60 \pm 0.30$ & $8.32 \pm 1.14$ & \cite{Richmond1996} & 1994I\\
    $28.90 \pm 0.69$ & $6.02 \pm 1.91$ & \cite{Baron1996} & 1994I\\
    $29.20 \pm 0.30$ & $6.29 \pm 0.96$ & \cite{Iwamoto1994} & 1994I\\
    \hline
    \multicolumn{4}{c}{\textbf{Planetary Nebula Luminosity Function (PNLF)}}\\
    $29.41 \pm 0.12$ & $7.62 \pm 0.42$ & \cite{Ciardullo2002} & \cite{Feldmeier1997}\\
    $29.52 \pm 0.12$ & $8.02 \pm 0.44$ & \cite{Ferrarese2000} & \cite{Feldmeier1997}\\
    $29.62 \pm 0.15$ & $8.40 \pm 0.58$ & \cite{Feldmeier1997} & \cite{Feldmeier1997}\\
    \hline
    \multicolumn{4}{c}{\textbf{Surface Brightness Fluctuations (SBF)}}\\
    $29.32 \pm 0.14$ & $7.31 \pm 0.47$ & \cite{Tully2013} & \cite{Tonry2001}\\
    $29.38 \pm 0.27$ & $7.52 \pm 0.93$ & \cite{Ciardullo2002} & \cite{Tonry2001}\\
    $29.42 \pm 0.27$ & $7.66 \pm 0.95$ & \cite{Tonry2001} & \cite{Tonry2001}\\
    $29.47 \pm 0.28$ & $7.83 \pm 1.01$ & \cite{Ferrarese2000} & \cite{Tonry2001}\\
    $29.59 \pm 0.15$ & $8.28 \pm 0.57$ & \cite{Richmond1996} & \cite{Richmond1996}\\
    \hline
    \end{tabular}
    }
    \caption{Previous distance measurements for M~51 from the literature, along with the new estimates. The list contains the data plotted in Fig.~\ref{fig:dists}. The Data column shows the source of the data that was utilised for the analysis, except for supernovae, in which case the individual objects are listed. For a more complete list of distances and a review on each method, see Table 2 in \cite{McQuinn2016}.}
    \label{tab:dists}
\end{table}

\begin{figure}
    \centering
    \includegraphics[width=\linewidth]{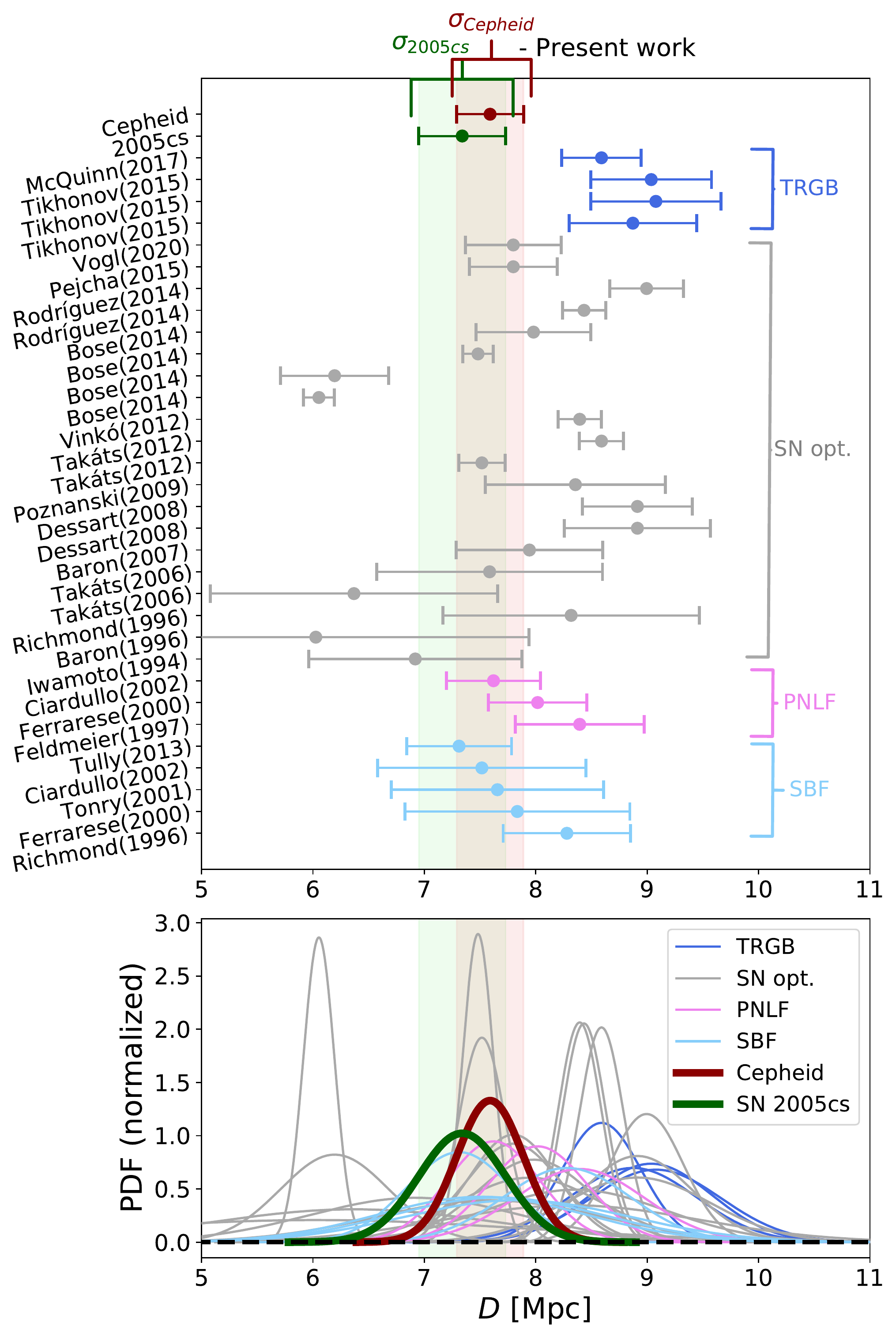}
    \caption{Comparison of the various distance estimates. \textbf{Top:} the individual distance estimates from multiple publications making use of the surface brightness fluctuations (SBF), the Tip of the Red Giant Branch (TRGB), or optical supernova observations. The coloured regions in the background show the $1\sigma$ ranges of estimates presented in this paper. \textbf{Bottom:} the individual distance probability distributions assuming Gaussian uncertainties.}
    \label{fig:dists}
\end{figure}

Fig.~\ref{fig:dists} shows the comparison of our distances with other, previous distance estimates. Given that we estimated the first Cepheid-based distance to M~51, there are no other results of the same method we can compare our value to. Comparing our SN~2005cs distance to the other SN II-based distances, we find that it is lower but not inconsistent with the previous EPM and Global Fitting estimates of \cite{Vogl2020} and \cite{Pejcha2015} ($D = 7.80 \pm 0.43$ and $D = 7.80 \pm 0.37$ respectively). Most of the difference between our results and theirs can be attributed to the different choices of time of explosion. Compared to the work presented in \cite{Vogl2020}, we applied a linear flux correction to the spectra, estimated the time of explosion based on light curve fits, used an extended set of model spectra for the emulator, and calculated the distance based on a different set of observed spectra. The tentative consistency of our EPM distance with this previous estimate shows that the uncertainties are realistic, i.e. the estimator is robust, despite the multiple changes made in the analysis. On the other hand, our EPM distance estimate is not in agreement with that calculated by \cite{Dessart2008}. We found that a significant portion of the mismatch can be explained by the 2.1-day difference in the adopted time of explosion. We note that the $t_0$ estimate of \cite{Dessart2008}, which is purely based on the EPM regression, is in tension with the last photometric non-detection of SN~2005cs. If we accounted for this time difference by fixing the time of explosion to the value obtained by \cite{Dessart2008}, the two results would be consistent.


The most relevant feature in Fig.~\ref{fig:dists} is the fact that both distances obtained by us are in at least $2\sigma$ tension with the TRGB values, even with the most recently obtained value from \cite{McQuinn2017} ($D = 8.58 \pm 0.44$ Mpc, updated from \citealt{McQuinn2016} to include systematic uncertainties). The difference between the distance estimates can be even more than 10\%, inducing a $\sim$ $0.2 - 0.4$ mag difference in the absolute magnitudes ($15-30$\% difference in flux). Since TRGB distances were used widely as a benchmark for studies which consider the absolute luminosities of objects within M~51 (for example, supergiant \citep{Jencson2022} or pulsar studies \cite{Brightman2022}), such a distance or luminosity difference influences the astrophysical results. It also causes a non-negligible change in the SNe IIP standardizable candle method as well, which uses SN~2005cs as a calibrator \citep{deJaeger2022}. Using the TRGB value, this object falls slightly away from the rest of the sample in terms of the calibrated absolute magnitude (see Fig.~1 of \citealt{deJaeger2022}). Assuming the newly estimated Cepheid distance, however, would put SN~2005cs closer to the rest of the sample, which in turn would revise the calibration parameters as well.

The explanation for such an offset between the TRGB and the Cepheid (or the EPM-based) distance is elusive. It cannot be ruled out that the offset is due to inherent systematic differences: for example, \cite{Anand2022} found an offset between the TRGB and the maser distance of NGC~4258, our anchor. However, the scale of this offset was only a few per cent, hence it cannot explain the entirety of the difference we found. There were also indications of offsets between Cepheid and TRGB distances, and consequently inferred Hubble constants, namely that the TRGB value of \cite{Freedman2021} ($69.8 \pm 2.2$ km s$^{-1}$) was found to be significantly different from the SH0ES estimate ($73.04 \pm 1.04$ km s$^{-1}$, \citealt{Riess2022}). The question, whether there is truly a systematic difference may be solved by acquiring a large enough sample of galaxies where both methods can be used for a precision distance estimation (see the $HST$ proposal of \citealt{2022hst..prop17079J}).


Another explanation for the offset may be that past TRGB results overestimate the distance of M~51. As discussed by \cite{Jang2021}, \cite{Freedman2021} and \cite{Madore2023}, the choice of the field (for example, if it is too close to the disk of the galaxy) can influence the analysis through the internal reddening of the host, or through blending. Moreover, \cite{Wu2022} and \cite{Scolnic2023} have shown that the apparent magnitude of the TRGB feature depends on the contrast between RGB and AGB stars near the tip. This may be important for the \cite{McQuinn2016} TRGB value, since the field chosen for analysis lies close to or partially even on the top of a spiral arm of M~51, hence the field is a lot more crowded, reddened and has a higher contribution from AGB stars than in usual TRGB analyses. On the other hand, the work conducted by \cite{Tikhonov2015} arrived at a distance that is consistent with the \cite{McQuinn2016} estimate, even though the underlying field was further away from M~51 (albeit \citealt{Tikhonov2015} used one of the earlier versions of the TRGB method). 

The method for determining the tip of the red giant branch on the CMD has been shown to influence the measurement, as shown by \cite{Wu2022}. Curiously, most of the inconsistency between our distances and the TRGB can be remedied if the absolute magnitude of the tip is assumed to be dimmer. Recent analyses (such as \citealt{Jang2021, Anand2022}) assume an absolute magnitude of about $M_{\textrm{F814W}} = -4.06$ mag (similarly to \citealt{McQuinn2016}). However, assuming a value of $M_{\textrm{F814W}} = -3.94$ mag, which is perfectly within the range of absolute magnitudes found by \cite{Rizzi2007}, would lead to a TRGB distance estimate of $\mu = 29.55$ mag. This is also supported by the recent analysis of \cite{Anderson2023}, who found a similarly fainter calibration magnitude for the TRGB method. This would be consistent with our estimates within $2 \sigma$. As noted by \cite{Rizzi2007}, the TRGB absolute magnitude weakly correlates with the metallicity even in the $I$ band, which could explain the distance offsets, given the super solar value valid for M~51.

It is worth noting that, despite the large offsets between our estimates and the TRGB value, the Cepheid and EPM distances are consistent with results of secondary distance indicators, such as PNLF ($7.62 \pm 0.42$ Mpc, \citealt{Ciardullo2002}) and SBF ($7.31 \pm 0.47$ Mpc, \citealt{Tully2013}). Naturally, these methods were calibrated based on Cepheids, hence a good agreement is expected (nevertheless it is important, that our estimates are independent of these, given that M~51 had no published Cepheid distance before our analysis, and the supernova distance requires no calibration). Based on these results, and the good agreement among all independent indicators except the TRGB, we find it more likely that M~51 is located closer to us than previously assumed.

\section{Conclusions}

The distance to M~51 was measured using two independent approaches: the well-established PL relation method of Cepheid variable stars, yielding a distance of $D = 7.59 \pm 0.30$ Mpc ($D = 7.49 \pm 0.30$ Mpc when both fundamental mode and overtone Cepheids are used) and by applying the tailored expanding photosphere method on SN~2005cs, resulting in $D = 7.34 \pm 0.39$ Mpc ($D = 6.92 \pm 0.69$ without using light curve information for the time of explosion estimation). Combining these two independent estimates yields a distance of $D_{\textrm{M~51}} = 7.50 \pm 0.24$ Mpc for M~51. The consistency of the obtained values demonstrates well the potential of SN IIP-based distances: even though the analysis does not rely on any calibration with other distance estimation methods, it produced a distance that is not only comparable in precision but is also in agreement with the Cepheid PL relation based result. A similar consistency was achieved between our results and some other secondary distance indicators, such as surface brightness fluctuations or the planetary nebula luminosity function.

Both of our estimates are in disagreement with the previously obtained TRGB values. It is unclear, whether this inconsistency is an inherent difference between the various methods, even though, former studies did not show systematic offsets of such magnitudes. Understanding this offset is important from the point of view of luminosity critical studies: since the difference between the newly obtained distances and the latest TRGB value is as large as 10\%, the choice of distance measure can significantly affect the astrophysical conclusions for objects within M~51.

This work also demonstrates that the improvements in the spectral modelling have put non-computation intensive and accurate Type IIP supernova distances well within reach. By obtaining spectral time series that are well suited for this type of analysis, one could use these supernovae to estimate distances well within the Hubble flow independently of the distance ladder. Such data had already been obtained by the Nearby Supernova Factory and the adh0cc collaborations, with the ultimate goal of inferring the local-Universe Hubble constant through tailored-EPM.

\section*{Data availability}
The data underlying this work is publicly available (via Wiserep for the \citealt{Pastorello2009} data, and via CDS for the \citealt{Conroy2018} catalogue). The data produced in this work, such as the final M~51 Cepheid catalogue and the flux calibrated spectral time series of SN~2005cs is available at the GitHub page of  the author: \url{https://github.com/Csogeza/M51}. The Cepheid catalogue will also be published at CDS.

\section*{Acknowledgements}

The authors thank Jason Spyromilio for the comments that helped improve the paper. The authors also thank Zoi Spetsieri for carrying out a check on stellar association bias for the Cepheid sample. The research was completed with the extensive use of Python, along with the \texttt{numpy} \citep{numpy}, \texttt{scipy} \citep{scipy} and \texttt{astropy} \citep{astropy} modules. This research made use of \textsc{Tardis}, a community-developed software package for spectral synthesis in supernovae \citep{TARDIS, TARDIS2}. The development of \textsc{Tardis} received support from the Google Summer of Code initiative and from ESA's Summer of Code in Space program. \textsc{Tardis} makes extensive use of Astropy and PyNE. CV and WH were supported for part of this work by the Excellence Cluster ORIGINS, which is funded by the Deutsche Forschungsgemeinschaft (DFG, German Research Foundation) under Germany's Excellence Strategy-EXC-2094-390783311. RIA acknowledges support from the European Research Council (ERC) under the European Union's Horizon 2020 research and innovation programme (Grant Agreement No. 947660). RIA further acknowledges funding by the Swiss National Science Foundation through an Eccellenza Professorial Fellowship (award PCEFP2\_194638). ST acknowledges funding from the European Research Council (ERC) under the European Union’s Horizon 2020 research and innovation program (LENSNOVA: grant agreement No 771776). SB thanks Sherry Suyu and her group at the Technische Universität München (TUM) for their hospitality. This work was supported by the `Programme National de Physique Stellaire' (PNPS) of CNRS/INSU co-funded by CEA and CNES.

\bibliographystyle{aa}
\bibliography{citations}

\newpage

\begin{appendix}
\section{Examples for differences between the Fourier method and GAM}
\label{sec:FG_comp}

\begin{figure}
    \centering
    \includegraphics[width = \linewidth]{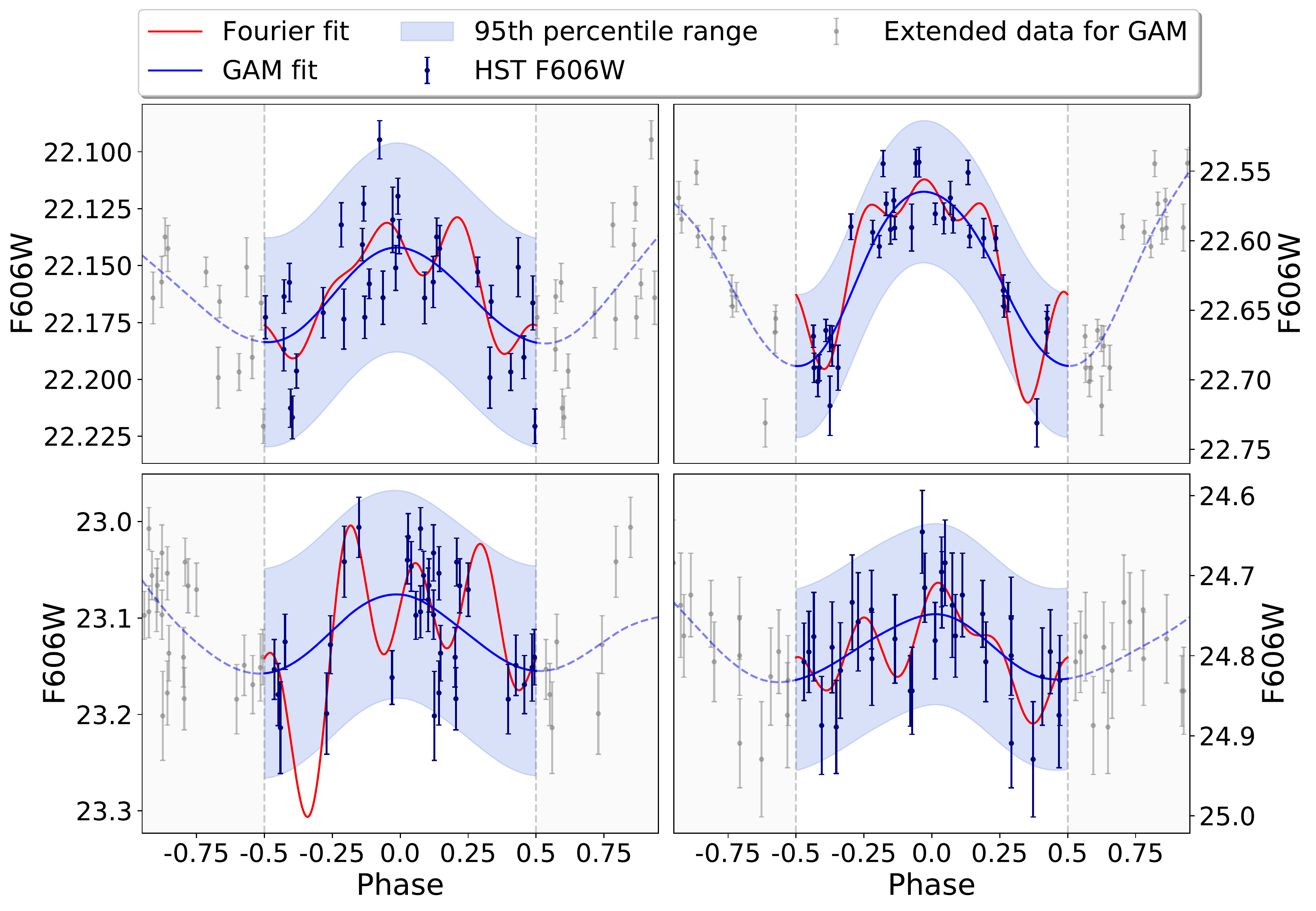}
    \caption{Example light curve fits, where the Fourier and the GAM method yield different results. Even though uncertainties are high, these variable stars were still included for the further filtering steps, owing to the smooth GAM fits, whereas the Fourier method would have removed using our approach.}
    \label{fig:app_comp}
\end{figure}

Fig.~\ref{fig:app_comp} shows examples of light curves, where the Fourier and GAM fits yielded significantly different results. In such cases, the naive Fourier method overfits the data. This can be avoided by limiting the number of Fourier terms used for the fitting, individually for each object. However, the GAM method performs well in these cases, resulting in smooth light curves, which can still be compared to the rest of the sample reasonably.

\section{Cepheid catalogue}
\label{sec:app}

In Tab~\ref{tab:appendix} the list of Cepheids found by our filtering is shown. This list is complete, i.e. even the stars that were flagged as outliers by the $\sigma$ clipping method are included. For the full table with photometric errors included, see the online version of the article.

\begin{table*}
\begin{center}
\begin{tabular}{|c c c c c c c c c|}
\hline
RAJ2000 & DEJ2000 & P& F555W & F606W & F814W & W$_{F555W, F814W}$ & [O/H] & $\sigma$-clip \\

[deg] & [deg] & [days] & [mag] & [mag] & [mag] & [mag] & [dex] & \\
\hline
\hline
202.516 & 47.171 & 51.81 & 23.350 & 23.087 & 22.384 & 21.190 & 9.160 & + \\
202.477 & 47.208 & 59.19 & 23.429 & 23.180 & 22.510 & 21.373 & 9.309 & + \\
202.496 & 47.177 & 37.60 & 23.420 & 23.179 & 22.525 & 21.419 & 9.240 & + \\
202.530 & 47.194 & 54.34 & 23.527 & 23.259 & 22.547 & 21.335 & 9.128 & + \\
202.530 & 47.184 & 48.93 & 23.380 & 23.143 & 22.495 & 21.400 & 9.124 & + \\
202.509 & 47.160 & 47.49 & 23.414 & 23.174 & 22.525 & 21.425 & 9.159 & + \\
202.460 & 47.184 & 45.04 & 23.623 & 23.325 & 22.551 & 21.227 & 9.317 & + \\
202.515 & 47.171 & 51.97 & 23.654 & 23.353 & 22.564 & 21.216 & 9.163 & + \\
202.516 & 47.171 & 41.07 & 23.445 & 23.212 & 22.573 & 21.495 & 9.160 & + \\
202.468 & 47.165 & 50.58 & 23.589 & 23.331 & 22.642 & 21.470 & 9.253 & + \\
202.531 & 47.180 & 42.14 & 23.866 & 23.521 & 22.649 & 21.145 & 9.117 & + \\
202.489 & 47.180 & 21.42 & 24.512 & 23.931 & 22.651 & 20.351 & 9.271 &   \\
202.517 & 47.165 & 38.41 & 23.668 & 23.403 & 22.700 & 21.503 & 9.144 & + \\
... & ... & ... & ... & ... & ... & ... & ... & ...  \\
\hline
\end{tabular}
\caption{List of Cepheid variables found in our analysis. This table includes all variables that remained in the sample after cutting away the stars with blue colour than the instability strip. The $\sigma$-clip column shows which variables were not classified as outliers by the $\sigma$ clipping algorithm we applied (denoted with the '+' symbol). For the full table, see the online appendix of the article. The metallicities are measured according to the \cite{Zaritsky94} scale.}
\label{tab:appendix}
\end{center}
\end{table*}

\end{appendix}

\end{document}